\documentclass[a4paper]{article}
\usepackage{times}
\usepackage{authblk}
\usepackage[]{algorithm, algorithmic}
\usepackage{graphicx}
\usepackage{amsmath, amssymb}
\usepackage{booktabs}
\usepackage{color}
\usepackage[usenames,dvipsnames]{xcolor}
\usepackage[round]{natbib}
\usepackage{appendix}
\usepackage{tcolorbox}
\usepackage{rotating}
\usepackage{multirow}
\usepackage{longtable}
\usepackage{todonotes}
\usepackage{array}

\tcbset{
  colback=white,
  colframe=black,
  arc=2mm,
  boxrule=0.7pt
}

\usepackage{orcidlink}

\providecommand{\keywords}[1]{\textbf{{Keywords---}} #1}

\begin{document}







\title{Adressing Separation: A Firth-corrected Joint Model for Longitudinal and Time-to-event Data with an Application on Dropout from Vocational Training}
\author[$1$, $2$,*]{Sophie Potts\orcidlink{0000-0003-0949-3887}}
\author[$3$]{Viola Deutscher\orcidlink{0000-0002-9714-6465}}
\author[$1$, $2$]{Elisabeth Bergherr\orcidlink{0000-0003-3983-9957}}

\affil[$1$]{ Chair of Spatial Data Science and Statistical Learning, Platz der Göttinger Sieben 3, 37073 Göttingen}
\affil[$2$]{ Campus Institute Data Science (CIDAS), Goldschmidtstraße 1, 37077 Göttingen}
\affil[$3$]{ Chair of Business Education --Digital Vocational Learning, Georg-August University Göttingen, Germany}

{\footnotesize{
\affil[*]{Corresponding author {\sf{e-mail: sophie.potts@uni-goettingen.de}}}}
\maketitle  

\begin{abstract}

\noindent Joint Models for longitudinal and time-to-event data are frequently used to model endogenous longitudinal covariates alongside a time-to-event outcome. However, the model class borrows some limitations of the survival submodels, including the necessity for non-separation for each category of categorical covariates. We therefore incorporate Firth's correction into the frequentist estimation procedure of joint models in order to make the model class applicable in settings with separation cases. We derive the needed quantities for the correction term and implement it in the Expectation-Maximization Algorithm for the parameter estimation in joint models. Our simulation study shows, that in data situations with separation issues, the Firth-corrected estimation procedure yields less biased estimates and the respective coefficients approach the estimated values observed in the non-separation cases. The application on a data set on satisfaction with and dropouts from vocational training demonstrates the advantages of the Firth-corrected joint model in a real world data set with separation. The results add to the literature on dropout from vocational training in Germany by explicitly modeling direct effects of socioeconomic and training-specific factors on the risk of dropout as well as their indirect contribution via satisfaction with the training.

\end{abstract}

\noindent \keywords{joint models, Firth correction, monotone likelihood, separation issues, rare events, vocational training dropout}

\section{Introduction}

\normalsize

When modelling a time-varying covariate alongside a time-to-event outcome the model class of joint models for longitudinal and time-to-event data \citep{FAUCETT1996, Wulfsohn1997} is often applied. It allows for proper modelling and incorporation of an endogenous covariate, such as a biomarker used for prediction of the risk of death. The model consists of two submodels which are estimated together: a longitudinal submodel and a survival submodel. While the longitudinal submodel employs a mixed model to account for the nested data structure of longitudinal observations, the survival model usually uses a proportional hazards model. 
 
As the joint model is constructed from the two submodels, some limitations of the survival model transfer directly to the joint model. For example in situations with separation, i.e.\ a certain category of a covariate does not exhibit any event, the model is inefficient or even fails to estimate a coefficient at all. This is due to the monotone likelihood, that arises from the separation. 
The problem of separation has been tackled in binary regression models \citep{firth1993} as well as in Cox models \citep{heinze2001} by using a correction term in the estimation procedure. In order to allow joint models to be applicable in situations of separation that are likely present in cases of of rare events and/or highly imbalanced covariates \citep{Bryson1981}, we transfer the correction to this model class. 


The rest of the paper is structured as follows: in Section 2, we introduce joint models formally. Subsequently, we describe the parameter estimation procedure via the Expectation-Maximization-Algorithm (EM) and showcase the limitations in separation settings. To overcome the limitations, our implementation of the Firth correction is explained in Section \ref{sec:FC} in more detail. In Section \ref{sec:sim}, we set up a simulation study to proof our concept and the results are presented. Section \ref{sec:data} yields a real world data application, in which we employ our corrected model to data on satisfaction with and dropout from vocational training in Germany where one professional sector yields a separation case. The results of the application are presented and the article concludes with a summary of the main results, limitations and an outlook on possible extensions. 




\section{Methods}

\subsection{Joint models for longitudinal and time-to-event data}
\label{sec:JM}

The model class of joint models for longitudinal and time-to-event data has been extensively used in biostatistics \citep[see e.g.\ ][]{Nez2014, Ferrer2016} and is transferred to other research areas as well \citep[e.g.\ social sciences:][]{Potts2026, Cremers.2021}. A joint model allows to incorporate an endogenous time-varying covariate in a classical time-to-event model. Accordingly, the endogenous covariate (e.g.\ a biomarker when looking at the risk of death in the time-to-event submodel or marital satisfaction when analysing the risk of marriage dissolution) is modelled via a linear mixed model (LMM), capturing intra-individual variance via random intercepts ($b_{0i}$) and random slopes ($b_{1i}$) for each individual $i=1, \ldots, N$. By incorporating the predictions of a longitudinal model in a survival model, the two models are linked. Note, that the estimation is performed simultaneously for the parameters of both sub-models, such that no two-stage approach is needed (see Section \ref{sec:EM}). In its simplest form, the model is defined as
\begin{equation}
    \begin{aligned}
y_i(t) &= m_i(t)+\varepsilon_i(t), ~~~~ & \varepsilon_i(t) \sim N(0, \sigma^2) \\
m_i(t) &= \boldsymbol \beta^T \boldsymbol{{x}}{_i}_{\text{long}}(t) +  \boldsymbol{b}_i^T\boldsymbol{{z}_i}(t), ~~~ & \boldsymbol{b_i} \sim N( \boldsymbol{0}, \boldsymbol{Q})\\
        h(t|M_i(t), \boldsymbol{x_i}) &= h_0(t) \exp [\boldsymbol \gamma^T \boldsymbol{{x}}{_i}_{\text{surv}} + \alpha~ m_i(t)], ~~~~ &
    \end{aligned}
    \label{eq:JM1}
\end{equation}
where $m_i(t)$ represents the true value of the longitudinal process at time $t$ and $M_i(t)$ depicts the whole trajectory of $m_i(t)$ until time $t$. The design matrix of the random effects $\boldsymbol{z_i}(t)$ captures the random intercept and slope on time $t$ per individual, such that $\boldsymbol{z_i}(t)= (1 ~~t)^T$ and $\boldsymbol b_i =(b_{0i} ~~ b_{1i})^T$. The fixed effects in the longitudinal submodel are denoted by $\boldsymbol{\beta}$ incorporating a fixed effect for the covariate time ($t$) as well. Besides classical exogenous covariates summarized in $\boldsymbol{{x}_i}_{\text{surv}}$, the survival submodel incorporates the value of the endogenous covariate via $m_i(t)$. It is equipped by a coefficient $\alpha$ which is called the \emph{association parameter}. In the following, the baseline hazard $h_0(t)$ is approximated via B-Splines involving the basis function coefficients $\boldsymbol \gamma_{bs}$ and a classical B-Spline basis function matrix.

Joint Models are especially useful when trajectories of time-varying covariates carry information about the time-to-event outcome, i.e.\ are endogenous as defined by \cite{Kalbfleisch2002}. They further allow to decompose effects of other covariates into direct effects on the time-to-event outcome and indirect effects via the endogenous covariate. The basic current value joint model as presented in Equation \eqref{eq:JM1} can be extended in various ways: multiple longitudinal trajectories linked to one survival outcome \citep[e.g.][]{Lin2002}, different types of outcome distributions \citep[e.g.][]{Viviani2014}, competing risks \citep[e.g.][]{Elashoff2008} and accelerated failure time models \citep[e.g.][]{Tseng2005} for the survival submodel as well as different association structures beyond the current value $m_i(t)$ \citep{rizbook}. For an overview of extensions see \cite{Papageorgiou2019}. In the following we concentrate on the joint model as defined in Equation \eqref{eq:JM1}, i.e.\ a current value association between one Gaussian distributed longitudinal outcome and a time-to-event model with one type of event. We further stick to time-constant covariates in the submodels, except for the covariate \emph{time} itself.

\subsection{Parameter estimation: Expectation-Maximization Algorithm}
\label{sec:EM}

The frequentist estimation of a joint model for longitudinal and time-to-event data is based on the Expectation-Maximization-Algorithm \citep[][Appendix B]{rizbook}. 
Treating the random effects in the longitudinal submodel as missing data enables the derivation of the expected value of the complete-data log-likelihood. This expectation is used in the maximization step to get the parameter estimates, that maximize $\mathbb E(\log(L))$ in this step. Iterating through the two steps results in an updating scheme for all parameters involved (see Box \ref{algobox}). In the case of a joint model, the parameters to estimate can be divided by the two submodels: The longitudinal submodel holds $\boldsymbol\theta_{\text{long}}=(\boldsymbol{\beta}^T, \sigma^2, q_{11},q_{12}, q_{22})$ and the time-to-event submodel contains $\boldsymbol \theta_{\text{surv}}=(\boldsymbol{\gamma}^T, \alpha, \boldsymbol{\gamma}_{bs}^T)$. The covariance parameters  $\boldsymbol{ \hat Q}$ and $\hat \sigma^2$ have a closed form solution \citep{rizbook} and the estimation of the fixed effects $\boldsymbol{\beta}$ is carried out by a one-step Newton update.

As separation cases interfere with the estimation of the coefficients in the survival submodel, we focus on their updating-scheme. Hereby $\boldsymbol{\hat \theta}_{\text{surv}}$ is updated in a Newton updating step via

\begin{equation*}
    \boldsymbol{\hat \theta}_{\text{surv}}^{(t+1)} 
    = \boldsymbol{\hat \theta}_{\text{surv}}^{(t)} 
      + \left[ I \left(\boldsymbol{\hat \theta}_{\text{surv}}^{(t)} \right)\right]^{-1}
        S \left( \boldsymbol{\hat \theta}_{\text{surv}}^{(t)} \right)    
\end{equation*}
where $S(\cdot)$ represents the score function and $I(\cdot)$ the Fisher Information matrix.

\begin{tcolorbox}
[title=Expectation-Maximization Algorithm for Joint Models, label=algobox]

\begin{tcolorbox}[title=E-step]
Compute the expected complete-data log-likelihood

\footnotesize
$$
\begin{aligned}
\mathbb{E} [ &\log L(\boldsymbol T, \boldsymbol \delta, \boldsymbol y, \boldsymbol b \mid \boldsymbol \theta) ] =
 \mathbb E \left (\sum_i \log \left (p(\underbrace{T_i, \delta_i}_{\substack{\text{survival} \\\text{outcome}}}, \underbrace{y_i}_{\substack{\text{longit.} \\\text{outcome}}}, \underbrace{b_i}_{\substack{\text{random} \\\text{effects}}}, \boldsymbol \theta) \right) \right ) \\
 = \sum_i &\int \left [ \log p(T_i, \delta_i |b_i; \boldsymbol \theta_{\text{surv}}, \boldsymbol{\beta}) + 
 \log p(\boldsymbol y_i | \boldsymbol b_i; \boldsymbol\theta_{\text{long}}) + \log p(\boldsymbol{b}_i ; \boldsymbol{Q}) \right] \times \\ 
 &p \left (b_i|T_i, \delta_i, y_i ; \boldsymbol \theta \right) db_i
 \end{aligned}
 $$

\normalsize
where the random effects are treated as missing data.
\end{tcolorbox}

\begin{tcolorbox}[title=M-step]
Maximize the expected complete-data log-likelihood
$$
\boldsymbol{\hat \theta}^{(t+1)} 
= \arg\max_{\boldsymbol \theta} \mathbb{E} \left [ \log L(\boldsymbol T, \boldsymbol \delta, \boldsymbol y, \boldsymbol b \mid \boldsymbol \theta) \right ] ~~~ \text{using}
$$

\begin{itemize}
    \item \textbf{Closed-form updates for:} $\sigma^{2\,(t+1)}
    \qquad
    \boldsymbol Q^{(t+1)}$

    \item \textbf{Newton–Raphson update for fixed effects } $\boldsymbol \beta$:
    \begin{tcolorbox}[colback=blue!5!white,colframe=blue!45!black]
    \[
    \boldsymbol{ \hat \beta^{(t+1)} }
    = \boldsymbol {\hat \beta^{(t)} } 
      + \left[ I \left(\boldsymbol {\hat \beta^{(t)} } \right)\right]^{-1}
        S \left( \boldsymbol {\hat \beta^{(t)} } \right),
    \]
    \end{tcolorbox}

    \item \textbf{Newton–Raphson update for time-to-event coefficients} $\boldsymbol{\gamma}, \alpha, \boldsymbol{\gamma}_{bs}$:
    \begin{tcolorbox}[colback=blue!5!white,colframe=blue!45!black]
    \[
    \boldsymbol{\hat \theta}_{\text{surv}}^{(t+1)} 
    = \boldsymbol{\hat \theta}_{\text{surv}}^{(t)} 
      + \left[ I \left(\boldsymbol{\hat \theta}_{\text{surv}}^{(t)} \right)\right]^{-1}
        S \left( \boldsymbol{\hat \theta}_{\text{surv}}^{(t)} \right),
    \]
    where 
    \(
    \boldsymbol{\hat \theta}_{\text{surv}}= ( \boldsymbol{\gamma}^T, \alpha, \boldsymbol \gamma_{bs}^T).
    \)
    \end{tcolorbox}

\end{itemize}
\end{tcolorbox}
\label{algo}
\end{tcolorbox}

\subsubsection{Estimation problems in monotone likelihood cases}
\label{sec:problem}

The parameter estimation procedure (presented in Section \ref{sec:EM}) exhibits problems in cases of separation, as the log-likelihood becomes monotone in the direction of the coefficient corresponding to the separating covariate, leading to diverging or infinite regression coefficients \citep{firth1993, heinze2001, Bryson1981}.
Exemplarily, consider a simple joint model with current value association as given in Equation \eqref{eq:JM1} with one imbalanced binary covariate $x_1$ and one metric covariate $x_2$. Both covariates are included in both submodels, such that $ \boldsymbol x_{\text{long}}=(\boldsymbol 1 ~ \boldsymbol x_l{_1}~ \boldsymbol x_l{_2} ~ \boldsymbol t) \text{~and~} \boldsymbol x_{\text{surv}} = (\boldsymbol x_s{_1}~ \boldsymbol x_s{_2})$. Whilst $\boldsymbol x_l{_1}$ and $\boldsymbol x_s{_1}$ contain identical information for each individual, the distinction between them indicates that the longitudinal submodel incorporates measurement points, leading to multiple rows per individual. The true parameter values are set to $\boldsymbol \beta = (2,-1, 1, 0.45)^T$, with the last value being $\beta_t$, $\gamma =(-2.5, 1)^T$ and $\alpha=1$ resulting in the distribution of events as shown in Table \ref{tab:separation}. 
\begin{table}[ht!]
    \centering
    \begin{tabular}{l|c|c}
         & censored  & event\\
         \midrule
         $x_1=0$& 283 & 43\\
         $x_1=1$ & 74 & 0
    \end{tabular}
    \caption{Separation issue regarding the number of events in an example joint model. N=400, maximum number of observation times: 9, $q_{11}=0.6^2, q_{12}=q_{21}=-0.0276, q_{22}=2.3^2$, $\sigma^2 =0.01^2$}.
    \label{tab:separation}
\end{table}
This separation leads to coefficient divergence for the covariate $x_1$ in the survival submodel. Using the joint model implementation of the R-package \texttt{JM} and estimating the baseline hazard with B-Splines (\texttt{method="spline-PH-GH"}) gives an estimate of $\hat \gamma_1 = -19.77$ (std. err. $4343.36$) after convergence.

\footnotesize{
\begin{verbatim}
Data Descriptives:
Longitudinal Process		           Event Process
Number of Observations: 3427	    Number of Events: 43 (10.8%)
Number of Groups: 400

Joint Model Summary:
Longitudinal Process: Linear mixed-effects model
Event Process: Relative risk model with spline-approximated
		baseline risk function
Parameterization: Time-dependent 

Variance Components:
             StdDev    Corr
(Intercept)  0.7093  (Intr)
time         2.3069 -0.1944
Residual     0.2434        

Coefficients:
Longitudinal Process
                  Value Std.Err  z-value p-value
(Intercept)      1.7892  0.0128 140.3054 <0.0001
as.factor(Xl1)1 -1.2584  0.0159 -78.9989 <0.0001
Xl2              0.9774  0.0032 304.7908 <0.0001
time             1.3947  0.0210  66.5422 <0.0001

Event Process
                   Value   Std.Err z-value p-value
as.factor(Xs1)1 -19.7701 4343.3642 -0.0046  0.9964
Xs2               1.0343    0.2106  4.9121 <0.0001
Assoct            0.9904    0.1531  6.4702 <0.0001
bs1              -9.2880    1.2286 -7.5598 <0.0001
bs2              -6.4973    1.4217 -4.5701 <0.0001
bs3              -8.9240    1.6016 -5.5718 <0.0001
bs4              -7.1165    1.0298 -6.9106 <0.0001

Integration:
method: Gauss-Hermite
quadrature points: 9 

Optimization:
Convergence: 0 
\end{verbatim}
}
%

\subsection{Firth-correction for Joint Models}
\label{sec:FC}

\normalsize
The coefficient estimation problem in cases of separation is well known and studied for models with a binomial-distributed outcome and time-to-event models where the logic of separation can easily be transferred to censored vs.\ non-censored observations (see Section \ref{sec:problem}). In data collection settings, where the number of events and its distribution is not a feature to be influenced, methods to tackle the issue of separation involve data engineering methods in a first place, such as collapsing groups, binning or dropping the respective covariate. However, in settings of research, where the effect of the highly stratifying covariate is of interest, non of these methods allows to obtain reasonable estimates. We therefore transfer and implement the penalization strategy of the Firth-correction into the joint model framework. Our implementation builds upon the routines available in the R package \texttt{JM} \citep{jmpackage}.
The main change compared to unpenalized EM estimation is a correction term in the score function of the survival coefficients.\footnote{To improve readability, we removed the index $\boldsymbol{\hat \theta}_{\text{surv}}$ in this section.}

Firth's correction  \citep{firth1993, heinze2001} can be used in cases of monotone likelihood, where the respective score function never approaches zero. The resulting difference between $S \left(\boldsymbol{\theta} \right)$ and $S \left(\boldsymbol{\hat \theta} \right)$ leads to infinite or insensible estimates and thus to a bias in $\boldsymbol{\hat \theta}$ \citep{firth1993}. The idea to eliminate that bias is to modify the score function by including a correction term, such that a root of the corrected score function exists and yields a bias-reduced estimate for $\boldsymbol{\theta}$. 
To derive the correction term, one first approximates the bias of $\boldsymbol{\hat \theta}$ via a second order Taylor expansion of the score function around the true value: 

\begin{equation*}
\label{eq:TE_score}
    S(\boldsymbol{\hat\theta}) \approx S(\boldsymbol{\theta}) + S'(\boldsymbol{\theta}) (\boldsymbol{\hat \theta} - \boldsymbol{\theta})+ \frac{1}{2}(\boldsymbol{\hat \theta} - \boldsymbol{\theta})^{T}S''(\boldsymbol{\theta})(\boldsymbol{\hat \theta} - \boldsymbol{\theta})
\end{equation*}
with $S'(\boldsymbol{\theta}) = -I(\boldsymbol{\theta})$.
Setting to zero, taking the expectation and rearranging yields

\begin{equation}
\label{eq:bias}
\begin{aligned}
    \text{bias}( \boldsymbol{\hat \theta}) =\mathbb E(\boldsymbol{\hat \theta} -\boldsymbol{\theta}) &\approx \frac{1}{2} I(\boldsymbol{\theta})^{-1} \mathbb E\left[ (\boldsymbol{\hat \theta} - \boldsymbol{\theta})^{T} S''(\boldsymbol{\theta}) (\boldsymbol{\hat \theta} -\boldsymbol{\theta}) \right] \\ 
    &\approx \frac{1}{2} I(\boldsymbol{\theta})^{-1} \text{tr} \left [ I(\boldsymbol{\theta})^{-1}  \frac{\partial I(\boldsymbol{\theta})}{\partial \boldsymbol{\theta}} \right].
\end{aligned}
\end{equation}
Further note, that the Maximum-Likelihood Estimate (MLE) is linear in the score as $\boldsymbol{\hat \theta} \approx \boldsymbol{\theta} + I(\boldsymbol{\theta})^{-1} S(\boldsymbol{\theta})$, i.e. shifting the score function by $c(\boldsymbol \theta)$, shifts the MLE by $I(\boldsymbol{\theta})^{-1}c(\boldsymbol{\theta})$. 
Since we seek for an unbiased estimate $\boldsymbol{\tilde \theta}$, we combine the derived bias from Equation \eqref{eq:bias} and the linear shift property of the MLE to solve Equation \eqref{eq:final_corterm} for $c(\boldsymbol{\theta})$ which results in the final correction term for the score function and eliminates the first order bias of the estimate $\boldsymbol{\hat \theta}$.

\begin{equation}
\label{eq:final_corterm}
    \begin{aligned}
        0 &\overset{!}{=} \mathbb{E}(\boldsymbol {\tilde \theta} - \boldsymbol \theta) \approx  \text{bias} (\boldsymbol{\hat \theta}) + I(\boldsymbol \theta)^{-1} c(\boldsymbol \theta) \\
        c(\boldsymbol \theta) &=-I(\boldsymbol \theta)~ \text{bias} (\boldsymbol{\hat \theta}) \\
        c(\boldsymbol \theta) &= -I(\boldsymbol \theta) \left(-\frac{1}{2} I(\boldsymbol\theta)^{-1}  \text{tr} \left [ I(\boldsymbol{\theta})^{-1}  \frac{\partial I(\boldsymbol{\theta})}{\partial \boldsymbol \theta} \right] \right) \\
        c(\boldsymbol \theta) &= \frac{1}{2}~ \text{tr} \left [ I(\boldsymbol{\theta})^{-1}  \frac{\partial I(\boldsymbol{\theta})}{\partial \boldsymbol \theta} \right] 
    \end{aligned}
\end{equation}

\noindent In order to overcome the estimation problems in cases of separation in joint models, we derive and implement the Firth-correction for the coefficient vector $\boldsymbol\theta_{\text{surv}} = (\boldsymbol \gamma^T, \alpha, \gamma_{\text{bs}}^T)$ by incorporating the correction term into the score function of the Newton-Raphson estimation in the maximization step for these coefficients. The corrected score function looks as follows: 
\begin{equation}
\label{eq: firth_score}
    S^*(\theta_r) = S(\theta_r) + c_r, ~~\text{with } c_r = \frac{1}{2}~ \text{tr} \left [ I(\boldsymbol{\theta})^{-1}  \frac{\partial I(\boldsymbol{\theta})}{\partial \theta_r} \right],    
\end{equation}
where $S(\theta_r)$ is the score function of the uncorrected model and $c_r$ indicates the correction term for the $r$th covariate in $\boldsymbol\theta_{\text{surv}}$. 
The derivation of the required terms and their respective formulae can be found in Appendix \ref{sec:AppendixA}.
Using these derived quantities allows to include the Firth-corrected score function $S^{*} \left( \boldsymbol{\hat \theta}^{(t)} \right)$ from Equation \eqref{eq: firth_score} into the estimation process, such that the last step in the EM-Algorithm in Box \ref{algobox} changes to 

\begin{tcolorbox}[colback=black!5!white,colframe=black!45!black]
\begin{itemize}
 \item \textbf{Newton–Raphson update for time-to-event coefficients} $\boldsymbol{\gamma}, \alpha, \boldsymbol{\gamma}_{bs}$:
 \end{itemize}
\begin{tcolorbox}
[colback=blue!5!white,colframe=blue!45!black]
    \[
    \begin{aligned}
    \boldsymbol{\hat \theta}_{\text{surv}}^{(t+1)} 
    &= \boldsymbol{\hat \theta}_{\text{surv}}^{(t)} 
      + \left[ I \left(\boldsymbol{\hat \theta}_{\text{surv}}^{(t)} \right)\right]^{-1}
        S^{*} \left( \boldsymbol{\hat \theta}_{\text{surv}}^{(t)} \right),
    \\
    \text{where~~~~~~~}
    \boldsymbol{\hat \theta}_{\text{surv}}&= ( \boldsymbol{\gamma}^T, \alpha, \boldsymbol \gamma_{bs}^T).
\end{aligned}
\]
\end{tcolorbox}
\end{tcolorbox}
\noindent We further change the starting values of the exogenous regression coefficients $\boldsymbol{\hat \gamma}$ to be the Penalized Maximum-Likelihood estimates from a classical Firth-corrected Cox model to speed up convergence.  For this, we rely on the R package \texttt{coxphf} \citep{coxphf}. 
The association parameter $\alpha$ is initiated using the estimate from a time-varying covariate approach in a Cox model and the initial values for the Spline coefficients are derived from fitting a Weibull model to the data, estimating the log-hazard of the model and using B-Splines to approximate this log-hazard, following the standard approach for joint models in \texttt{JM}. 

In contrast to \texttt{JM} which replaces the analytical Newton-Raphson step by a BFGS optimization and approximates the Hessian (and by this the Fisher Information matrix $I \left (\boldsymbol{\hat \theta}_{\text{surv}}^{(t)} \right)$, we rely on the analytical Hessians. 

Regarding inference, we follow the procedure of \cite{rizbook} and use the forward difference approximation of the full Hessian to derive the standard errors
\begin{equation*}
    \text{se}(\boldsymbol{\hat \theta}) = \sqrt{\text{Var}(\boldsymbol{\hat \theta})} = \sqrt{\text{diag} \left( \left(-H \left(\boldsymbol{\hat \theta} \right) \right)^{-1} \right)}.
\end{equation*}
We implemented this Firth-corrected estimation routine for a joint model with time-constant exogenous covariates, a current-value association and a B-Spline baseline hazard approximation. The R-Code for the algorithm, as well as a short demonstration, can be found in the GitHub repository \texttt{[https://github.com/SophiePotts/JM-FC]}.


\section{Simulations}
\label{sec:sim}
\subsection{Setup}

We conduct a simulation study with a special focus on separation cases to evaluate the Firth-corrected algorithm presented above. The setup  includes a joint model with two covariates that are part of both submodels simultaneously..
We varied the \textit{total share of events} in the data set across three levels $\%e \in \{45, 25, 10 \}$. The value $\%e=45$ corresponds to the baseline condition without additional censoring. For the remaining levels, additional censoring was introduced. This censoring was drawn from an exponential distribution. Furthermore, we varied the \textit{number of observations} $N \in \{50, 250\}$.
Both parameters ($N$ and $\%e$) were varied to test the Firth-corrected model under different degrees of information. 
Note that a fixed event proportion at a larger sample size $N$ corresponds to a larger absolute number of events. Consequently, the setting $N=50, \%e=10$ contains less information than the setting $N=250, \%e=10$. In general, the likelihood exhibits higher curvature when the absolute number of observed events increases, for example in the setting $N=250$ and $\%e=45$.

The aforementioned choices, alongside the maximum number of longitudinal measurements per individual, $n=9$, determine the maximum total number of observation time points.
The actual size of the longitudinal data set depends on the censoring and event times and is indicated by $D$. In the longitudinal model time is defined as $t \in [0,1]$. Measurement times for each individual $i$ are evenly spaced over the domain of $t$. Other parameters were kept constant, such that our model results in

\begin{equation}
\begin{aligned}
    y_i(t) &= \underbrace{2.5 + 1 x_{i1} - 1 x_{i2} + 0.95 t + b_{i0} + b_{i1}t}_{m_i(t)} +\varepsilon_i \\
    h(t|\mathcal{M}_i(t), \boldsymbol{x}) &= h_0(t) \exp(-3.5 x_{i1} + 0.5 x_{i2} + 0.5 m_i(t)),
\end{aligned}
\end{equation}
 for $i=1,\ldots, N$ with $\varepsilon_i \sim \mathcal{N}(0, 0.01^2)$ and $\boldsymbol{b}_i\sim \text{MVN} \left(\boldsymbol{0}, \left( \begin{array}{cc}
    0.6^2 & -0.0156 \\
     -0.0156& 1.3^2
\end{array} \right ) \right )$. \vspace{0.1cm} 

\noindent Values for the dichotomous variable $x_1$ were drawn from a binomial distribution with parameter $p=0.15$ to ensure an imbalanced covariate, that may result in separation cases.
Observations for the continuous covariate $x_2$ were drawn from a Gaussian distribution $\mathcal{N}(0,4)$. The survival times are simulated from a Weibull distribution, implying the baseline hazard $h_0(t) = \exp(\phi) \times k \times t ^{k -1}$. In our setting $\phi=-1.75,~ k=1.6$. 
For each setting we simulate 100 runs without separation and 100 runs resulting in separation cases.




\subsection{Results}
\label{sec:sim_results}



To ensure 100 valid runs per setting and separation status, 150 simulation runs were performed for each configuration. Runs were considered invalid if either the classical joint model (JM) or the Firth-corrected joint model (FC-JM) estimation failed and returned an error. The following results are based on 100 drawn simulation runs per setting and separation status. We compare the results of the classical joint model, the classical joint model with Firth-corrected starting values and our Firth-corrected joint model.

Figure \ref{fig:sim_study_coef1} shows boxplots of the estimated coefficient $\hat \gamma_1$ from the three algorithms for the potentially separating covariate $x_1$. The true value is indicated by the horizontal dashed line.
The overall trend shows the superior performance of the Firth-corrected joint model (FC-JM) in separation cases. The bias reduction relative to the baseline joint model (JM) and JM with Firth-corrected starting values (JMc) is substantial in all tested settings with separation. This is independent of the sample size $N$ and the event share. FC-JM pulls the estimate towards the estimates of the models in the non-separation cases. As expected, in non-separation cases, FC-JM exhibits a slightly greater average bias compared to the classical JM: In cases where the likelihood is not monotone, the correction incorporates additional curvature information that biases the estimates towards zero.

\begin{figure}[H]
    \centering
    \includegraphics[width=\linewidth]{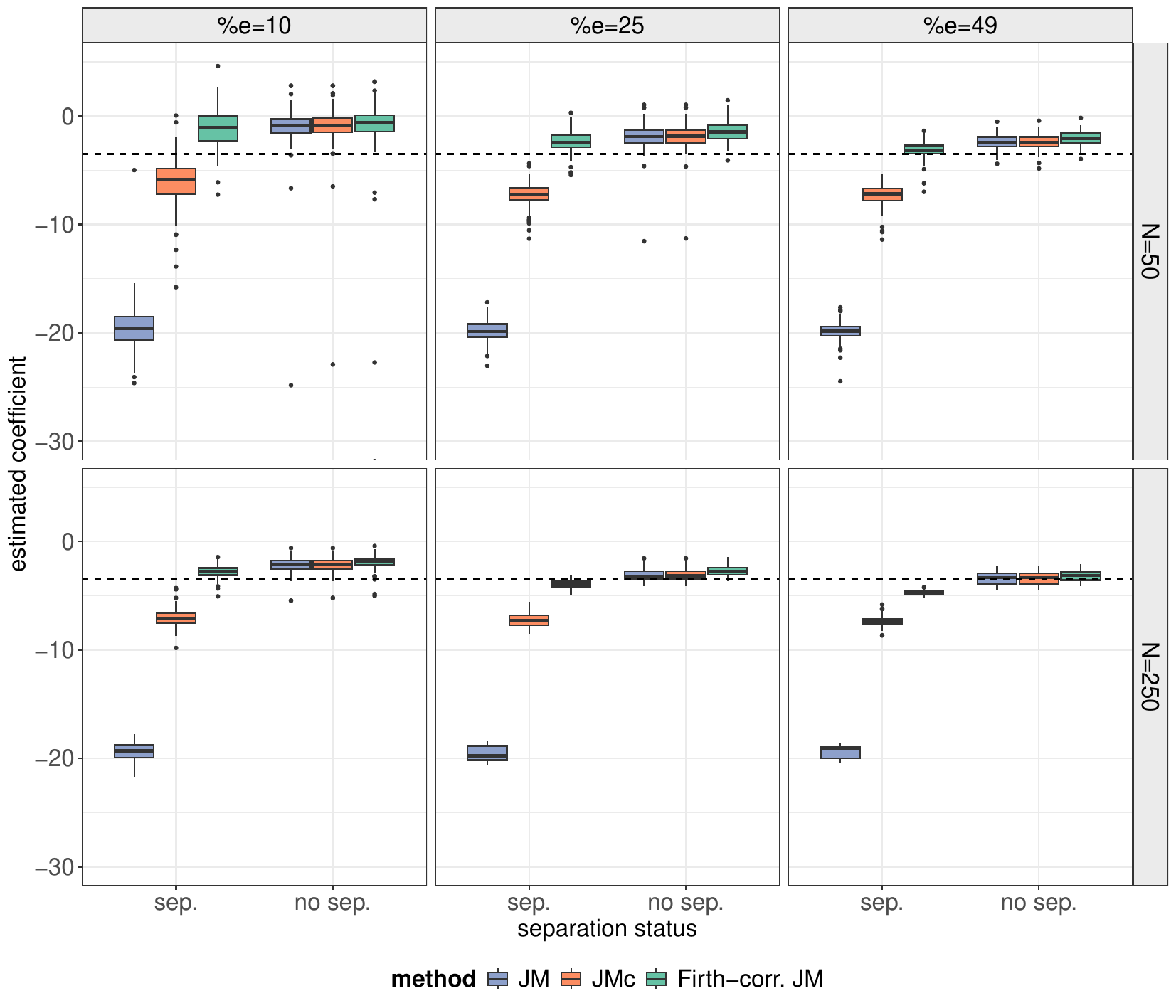}
    \caption{Results of the simulation study. Coefficient estimates $\hat \gamma_1$ for the potentially separating variable $x_1$ by number of individuals $N$, average share of events $\%e$ and separation condition from 100 simulation runs per setting and separation condition.}
    \label{fig:sim_study_coef1}
\end{figure}

The simulation study shows that the performance of FC-JM depends on the curvature of the likelihood, with the correction performing best in less informative settings such as N=50 and \%e=10.\footnote{The “best performance” is defined relative to the non-separation estimates rather than the true coefficient value.} The results indicate, that the performance of the correction is closely related to the raw number of events: In cases with fewer absolute events and separation, FC-JM performs more similarly to the corresponding non-separation scenarios, whereas larger numbers of events are associated with increased bias under separation. At the same time, very small numbers of events are also associated with an overall higher bias of the coefficient estimates, although this effect occurs regardless of separation status.


A small number of outliers (7/1,200) with estimates of $|\hat \gamma_1| > 30$ for FC-JM  are not depicted in Figure \ref{fig:sim_study_coef1}. They are attributable to ill-conditioned analytical Hessians that appear most likely in settings with very few information.These do not arise in the baseline JM, as this method approximates the Hessian via BFGS and yields more numerically stable matrices.


Regarding the other estimates of the joint model for longitudinal and time-to-event data, the median absolute bias for each coefficient in the model within the different settings of the simulation study can be found in Appendix Table \ref{tab:sim_res}. The results indicate, that the estimation accuracy among the other coefficients is basically identical in corrected and uncorrected models. 
As FC-JM outperforms JM in separation cases, wepromote its use in these specific contexts. The degree of curvature of the likelihood, driven by the raw number of events, is the key factor governing how well the correction performs relative to the non-separation baseline models.

\section{Data application: Dropout from vocational training}
\label{sec:data}

We apply the Firth-corrected joint model to a real-world data set on dropouts from vocational training in Germany. Leaving vocational training at own request can have unfavorable individual consequences for the apprentices and is costly for the training firms as well \citep{bohn2022dropout, michaelis2024long}. Therefore, a growing body of research investigates the impact of potential predictors of dropout, e.g.\ with a focus on informedness \citep{Herrmann2024}, career compromise in terms of field of work, social status, and gender share \citep{Beckmann2023}, previous educational background \citep{michaelis2022influence, ma2024exploring}, or the perceived quality of the training \citep{Krotz2021}. In the studies, satisfaction with training and dropout are either modeled as separate outcomes in independent analyses \citep{volodina2015success, michaelis2022influence}, or satisfaction is included as a predictor in discrete-time event history models of dropout, without accounting for its longitudinal development \citep{Siembab2023}. In both approaches, the mediating role of satisfaction remains formally unexamined. The existing literature therefore does not decompose covariate effects into direct effects on the dropout risk and indirect effects operating via the satisfaction pathway. To address this gap, we employ a joint model for longitudinal and time-to-event data, which simultaneously estimates the trajectory of training satisfaction and the hazard of self-initiated dropout, thereby enabling a formal test of the mediating role of satisfaction. To the best of our knowledge, this modeling approach has not yet been applied in the field of dropout from vocational training.

Theoretically, we justify this decision by the observation, that the decision to leave vocational training prematurely is rarely an abrupt and single cause event but rather the endpoint of a multi-cause cumulative process of dissatisfaction and disengagement over time \citep{bohn2022dropout}. This process perspective provides the theoretical foundation for treating satisfaction as a mediator rather than a parallel outcome or a static predictor of dropout. Structural and individual characteristics --- such as career compromise, educational background, or training quality --- shape the day-to-day experience of the apprenticeship and thereby influence satisfaction with the training on an ongoing basis. Dissatisfaction, in turn, increases the subjective cost of remaining in the training relationship and lowers the threshold for contract termination \citep{michaelis2022influence}. Treating satisfaction as a parallel outcome would leave this mechanism implicit and preclude any decomposition of covariate effects into a part that operates through the affective experience of training and a part that directly affects the dropout hazard regardless of satisfaction levels. Treating satisfaction merely as a time-constant predictor, on the other hand, ignores the dynamic nature of the construct: satisfaction evolves over the course of training, and it is precisely this trajectory --- rather than a single baseline measurement --- that is theoretically expected to govern dropout risk at any given point in time. A joint model for longitudinal and time-to-event data directly operationalizes this causal chain: the longitudinal submodel captures the time-varying satisfaction process as a function of covariates, while the survival submodel links the current value of that process to the instantaneous dropout hazard, net of any direct covariate effects. This specification is therefore not only a statistical convenience but a formal translation of the theoretical claim that satisfaction mediates the relationship between structural predictors and the risk of leaving vocational training.

\subsection{Data set}

We use German National Education Panel Study (NEPS) data from Starting Cohort 4 \citep{NEPS} to estimate the impact of drivers of dropout from vocational training at own request. 
For the pre-processing, we rely on the open source material from \cite{Beckmann_code} but use a more recent version of the data set (from 2010-2023) \footnote{NEPS SC4 Scientific Use File, version 16.0.0.} and include the time-varying variable on satisfaction with the vocational training.  The event variable comprises the timing of dropout at own request and dropout in agreement between employer and apprentice. In line with \cite{Beckmann2023}, dismissals are coded as censored observations. Time was measured in months since the start of the first vocational training and subsequently rescaled to years without loss of information.

Using the complete cases ($N = 4,587$) as a starting point and further restricting the sample to respondents with at least one measurement of satisfaction, our final data set includes $N = 4,203$ individuals (2,071 women; 2,132 men).
We are aware that this routine excludes especially individuals that have an early event (i.e. dropout early from their vocational training) as they rarely have any satisfaction measurement. Thus, the survival curve is systematically more adverse than the one of the retained sample, as the Kaplan--Meier comparison in Appendix \ref{sec:AppendixC} illustrates. This implies that our estimates will be conservative with respect to the strength of the satisfaction--dropout association.  Furthermore, this procedure results in a slight bias towards a higher average level of education as well as less respondents with migrational background in our final sample. A comparison of the complete case data set and our final sample can be found in Appendix \ref{sec:AppendixC} \footnote{Our results are not intended to replicate the findings of \cite{Beckmann2023} but showcase the usefulness of joint models on a data set of dropouts from vocational training as well as the performance improvement of the Firth- corrected joint model.}. 
Building on the literature discussed in Section \ref{sec:data}, we include a common set of covariates in both submodels and conduct separate analyses for men and women. The covariate \textit{sector of vocational training} leads to a separation case in the model for women (see Table \ref{tab:sep_neps})

\begin{table}[htb!]
    \centering
    \begin{tabular}{p{4.9cm}|cc|cc}
    & \multicolumn{2}{c}{Women} & \multicolumn{2}{c}{Men} \\
       Sector  & Censored & Event & Censored & Event \\
       \hline
       Production of goods  & 172& 13&1139 & 69\\
        Personal services& 967& 83& 221&37\\
        Business administration \& other business related services & 722& 35& 381& 21\\
        Services in IT and natural sciences &44 &1 &125 &11\\
         Other occ.\ in commercial services
& 34& 0& 123&5 \\
        
    \end{tabular}
    \caption{Distribution of respondents in sectors of vocational training by event status (dropout) and sex.}
    \label{tab:sep_neps}
\end{table}

\subsection{Model and Results}

Using the classical joint model, the separation caused by the covariate of occupational sectors results in an overly large regression coefficient and an inflated standard error (women: $\hat \gamma_{\text{sector5}}=-13.5608$,  $\text{se}=1748.1488$) (see Appendix \ref{append: jm_vs_fcjm}). This indicates the need for a correction. We hence use the Firth-corrected joint model to get properly estimated parameters and inference. 
We use a current value association, as well as a B-spline approximation for the baseline hazard. The longitudinal model incorporates a random intercept and a random slope on time:
\begin{equation*}
\begin{aligned}
    m_i(t) &= \beta_0+ b_{0i} +\boldsymbol \beta^T \boldsymbol x_{i} + \beta_t {t}  + b_{1i} {t} \\
    h(t|M_i(t), \boldsymbol{x}_i)&= \boldsymbol{\gamma}_{\text{bs}}^T\boldsymbol{x}_{\text{spline}}(t) \times \exp [\boldsymbol{\gamma}^T \boldsymbol{x}_i + \alpha ~m_i(t)].
\end{aligned}
\end{equation*}
The covariate vector $\boldsymbol{x}_i$ for person $i$ includes the following covariates: parental socioeconomic status, Grade Point Average (GPA), region of residence, type of training, degree of compromise in the field of work, migration background, highest educational degree and sector of vocational training. The definitions of the covariates have been taken from \cite{Beckmann2023}.
In the following we will present the analysis for women (see Table \ref{tab: res women}).

\begin{table}[hbt!]
\caption{Results for \emph{women}: Regression coefficients of the Firth-corrected joint model for longitudinal and time-to-event data.}
\label{tab: res women}
\resizebox{\textwidth}{!}{
\small
\centering
\begin{tabular}{p{6cm}rrr|rrr}
\toprule
       & \multicolumn{3}{c}{Longitudinal submodel} &  \multicolumn{3}{c}{Time-to-event submodel} \\
\vspace{0.1cm}
     & \multicolumn{3}{c}{(Satisfaction with vocational training)} &  \multicolumn{3}{c}{(Risk of dropout)} \\
\vspace{0.1cm}
  Variable & Estimate & Std.\ err. & p-value & Estimate & Std.\ err. & p-value \\ 
\midrule
(Intercept) & 8.5135 & 0.2238 & 0.0000 & & & \\
Time & -0.4022 & 0.0249 & 0.0000 & & & \\
Parental socioeconomic status (ISEI) & -0.1072 & 0.0322 & 0.0008 & 0.1255 & 0.0909 & 0.1641 \\
GPA & -0.0519 & 0.0305 & 0.0869 & 0.1065 & 0.0874 & 0.2185 \\
Residence region: East Germany $^a$ & -0.0617 & 0.0878 & 0.4728 & -0.2495 & 0.2662 & 0.3417 \\
Type of training: firm-based $^b$ & 0.0274 & 0.0739 & 0.6969 & -0.0476 & 0.1998 & 0.7956 \\
Field of work: weak compromise $^c$ & -0.1246 & 0.0833 & 0.1318 & 1.0607 & 0.2435 & 0.0000 \\
Field of work: moderate compromise $^c$ & -0.1811 & 0.0911 & 0.0459 & 0.5206 & 0.2868 & 0.0681 \\
Field of work: strong compromise $^c$ & -0.2883 & 0.0785 & 0.0002 & 0.7909 & 0.2450 & 0.0012 \\
Migration background: first generation $^d$ & -0.0335 & 0.1484 & 0.8051 & 0.4498 & 0.3419 & 0.1845 \\
Migration background: second generation $^d$ & 0.0070 & 0.0822 & 0.9133 & -0.2062 & 0.2526 & 0.4061 \\
Education: intermediate secondary $^e$ & -0.0817 & 0.0858 & 0.3339 & -0.4392 & 0.2162 & 0.0414 \\
Education: higher secondary $^e$ & -0.1800 & 0.0938 & 0.0538 & -1.2642 & 0.2634 & 0.0000 \\
Sector: production of goods $^f$ & 0.2549 & 0.2217 & 0.2452 & 1.0242 & 0.8968 & 0.2484 \\
Sector: personal services $^f$ & 0.2093 & 0.2041 & 0.2991 & 1.1657 & 0.8625 & 0.1730 \\
Sector: business administration \& other business related services $^f$ & 0.1412 & 0.2050 & 0.4811 & 0.6423 & 0.8697 & 0.4510 \\
Sector: other occ. in commercial services $^\text{f}$ & 0.5007 & 0.3076 & 0.1016 & -0.1469 & 1.6393 & 0.9100 \\
Satisfaction ($\hat \alpha$) & & & & -0.7114 & 0.0933 & 0.0000 \\
\bottomrule
\end{tabular}
}
\small{Reference categories:
$^a$ West Germany, 
$^b$ school-based,
$^c$ no compromise,
$^d$ no migration background,
$^e$ lower secondary,
$^\text{f}$ Sector: services in IT and natural sciences}
\end{table}
\noindent As expected, the model states a negative association parameter ($\hat\alpha = -0.7114$). The higher the satisfaction with the vocational training the lower the risk of dropout. This relationship is statistically significant and confirms that satisfaction is not merely correlated with dropout but continuously governs dropout risk throughout the training period, consistent with process-based and dynamic views of dropout and professional development \citep{bohn2022dropout, michaelis2022influence, deutscher2026training}. The covariates \textit{parental socioeconomic status} ($\hat \beta = -0.1072$), \textit{GPA} ($\hat \beta = -0.0519$), \textit{field of work compromise} and \textit{education} contribute significantly to the explanation of (dis)satisfaction. Especially, the strength of the compromise of the field of work has an impact on the satisfaction. On average the satisfaction with the vocational training decreases slightly yet significantly with \textit{time} ($\hat \beta = -0.0335$).

In the time-to-event submodel we however observe covariates that exhibit a direct effect on the risk of dropout even when controlled for the satisfaction level of the apprentice. This applies to \textit{education} (intermediate (higher) secondary vs.\ lower secondary: $\hat \gamma = -0.4392$ (-$1.2642$)) and \textit{field of work compromise} (e.g.\ strong compromise vs.\ no compromise $\hat  \gamma = 0.7909$). The residual positive direct effect of field of work compromise on self-initiated dropout may underline the stop-out interpretation proposed by \citet{Holtmann2023}, whereby apprentices use a deliberate contract termination to change the educational track after the leave of the first vocational training, irrespective of how satisfied they currently feel about the training. 
The finding that apprentices with higher educational backgrounds face a reduced dropout risk replicates evidence from prior longitudinal research \citep{ma2024exploring}. Crucially, the FC-JM extends this result by showing that the protective effect persists after conditioning on the full satisfaction trajectory, indicating that education does not operate solely through differences in affective experience. The residual gradient therefore likely reflects additional mechanisms, such as better pre-entry informedness, differential exit options, or selection into more prestigious vocational programs \citep{Beicht2014}.

The joint model also reveals that parental socioeconomic status, and region of residence show no significant direct effects on dropout hazard in neither the model for women nor for men. These characteristics do mainly shape the satisfaction trajectory, such that their total effect on dropout operates primarily through the affective experience of training rather than through structural constraints.

Furthermore, the effect of the sector on other occupations in commercial services (e.g.\ security, transportation and logistics, cleaning) causing separation is now properly estimated using the Firth-correction and indicates that there is no significantly lower risk of dropping out compared to the reference sector ($\hat \gamma = -0.1469,~ \text{p-value}=0.91$).

A comparison with the uncorrected model estimated using \texttt{JM}, as well as the corresponding results for men, are provided in Appendix \ref{append: jm_vs_fcjm}. The comparison with the uncorrected model demonstrates that the correction not only substantially affects the estimate of the separating covariate, but also alters other coefficients in the survival submodel. These differences are particularly pronounced for covariates associated with a small absolute number of events. For example, the coefficient for \emph{Sector: production of goods} changes from $\hat \gamma_{\text{uncor}} = 1.8234$ to $\hat \gamma_{\text{cor}} = 1.0242$, while the estimate for \emph{Migration background: first generation} changes from $\hat \gamma_{\text{uncor}} = 0.4048$ to $\hat \gamma_{\text{cor}} = 0.4498$.

\section{Conclusion}

In this work we implemented a modified version of the estimation of joint models for longitudinal and time-to-event data which accounts for the optimization problems in data with separation by extending the algorithm with a Firth-type correction.
The proposed estimation routine allows to estimate finite, less biased model coefficients in cases of separation in the time-to-event submodel.
The simulation study confirmed that the Firth-corrected joint model (FC-JM) substantially reduces bias for the separating covariate across all tested combinations of sample size and event share, while leaving estimates for all other coefficients essentially unchanged. 
FC-JM largely closes the gap between separation and non-separation scenarios, though the extent to which it does so is determined primarily by the degree of information in the likelihood: in low-information settings (i.e.\ high censoring, small $N$) it performs comparably to the non-separation baseline, whereas higher event shares are associated with residual bias under separation. A small number of numerical instabilities arising from ill-conditioned analytical Hessians point to a practical limitation of the approach and motivate further work on stabilization strategies for very sparse data.


At the level of content, the joint model on satisfaction with and dropout from vocational training enhances interpretive precision by placing established findings within a formally specified, simultaneous, and continuous causal logic. In doing so, it both replicates and extends the existing literature.}
The application highlights the usefulness of a joint model whenever a time-varying affective state partially mediates the influence of structural predictors on a time-to-event outcome --- a constellation that is likely to arise across sociology and educational research wherever subjective dispositions --- such as satisfaction, identification, or commitment --- are theorized as both products of structural conditions and proximate drivers of behavioral outcomes. 
The proposed Firth-corrected joint model offers a viable estimation strategy in settings where classical approaches fail due to monotone likelihood arising from rare events or strongly imbalanced categorical covariates.

However, several limitations remain. The current implementation is restricted to a current-value association structure and time-constant exogenous covariates. Extensions to current-slope associations, multiple longitudinal endogenous covariates, or \emph{time-varying} exogenous covariates \citep{HeinzeDunkler2008} would increase the flexibility of the model and are a natural direction for future work in both directions - methodological and regarding the application. Further, the standard errors rely on a forward-difference approximation of the full Hessian and should be interpreted with appropriate caution, particularly in the sparse-data settings where the correction is most needed.


\subsection*{Acknowledgements:} The work on this article was supported by the DFG (Deutsche Forschungsgemeinschaft; Projekt WA 4249/2-1). We further acknowledge support by the Open Access Publication Funds of the Göttingen University. 
This paper uses data from the National Educational Panel Study (\cite{NEPS}; \cite{Blossfeld}). The NEPS is carried out by the Leibniz Institute for Educational Trajectories (LIfBi, Germany) in cooperation with a nationwide network.

\subsection*{Author Contribution Statement:}
\textbf{Sophie Potts:} Conceptualization, formal analysis, methodology, software, data analysis, validation, visualization, writing -- original draft preparation, writing -- review \& editing. \textbf{Viola Deutscher:} Interpretation of dropout results, theoretical contextualization, writing -- review \& editing. \textbf{Elisabeth Bergherr:} Conceptualization, methodology, writing -- review \& editing.
\newpage
\bibliographystyle{apalike}  
\bibliography{bib}

\newpage
\appendix 

\setcounter{table}{0}
\renewcommand{\thetable}{\Alph{section}.\arabic{table}}

\setcounter{figure}{0}
\renewcommand{\thefigure}{\Alph{section}.\arabic{figure}}

\setcounter{equation}{0}
\renewcommand{\theequation}{\Alph{section}.\arabic{equation}}

\section{Components of Firth Correction for a Joint Model}
\label{sec:AppendixA}

{\small

We will here explain the details of the components used in the Expectation-Maximization Algorithm for Firth-corrected joint models. We focus on the Firth-corrected score function, which has the form

\begin{equation*}
    S^*(\theta_r) = S(\theta_r) + c_r,
    \qquad
    c_r = \frac{1}{2}\,\mathrm{tr}\left[I(\boldsymbol{\theta})^{-1}
          \frac{\partial I(\boldsymbol{\theta})}{\partial \theta_r}\right].
\end{equation*}

\noindent The components of the score functions are derived directly from the log-likelihood of a joint model with a current-value association structure, time-constant exogenous covariates in the survival sub-model, random intercept and random slope on time in the longitudinal sub-model, and a baseline hazard approximated by B-splines.

\subsection*{Log-likelihood}

The log-likelihood of a joint model consists of the three parts stemming from: the survival submodel, the longitudinal submodel and the distribution of the random effects $\boldsymbol{b}\sim \mathcal{N}(\boldsymbol 0, \boldsymbol{Q})$.

\begin{equation*}
\begin{aligned}
    \ell(\boldsymbol{\theta}\mid\boldsymbol{T},\boldsymbol{\delta},\boldsymbol{y},\boldsymbol{b})
    &= \sum_i \log \left[
         \Bigl(h_0(T_i)\exp\bigl(\boldsymbol{\gamma}^{\top}\boldsymbol{x}_{i}
               +\alpha\,m_i(T_i)\bigr)\Bigr)^{\delta_i}
         \times\exp \left(-\int_0^{T_i}h_0(s)
               \exp \bigl(\boldsymbol{\gamma}^{\top}\boldsymbol{x}_{i}
               +\alpha\,m_i(s)\bigr)\,ds\right)
       \right] \\[4pt]
    &\quad+\log \left[
         \bigl(2\pi\sigma^2\bigr)^{-n_i/2}
         \exp \left(-\frac{\|\boldsymbol{y}_i
               -\boldsymbol{\beta}^{\top}\boldsymbol{x}_i
               -\boldsymbol{b}^{\top}\boldsymbol{z}_i\|^2}{2\sigma^2}\right)
       \right] \\[4pt]
    &\quad+\log \left[
         (2\pi)^{-q/2}\det(\boldsymbol{Q})^{-1/2}
         \exp \left(-\tfrac{1}{2}\boldsymbol{b}_i^{\top}\boldsymbol Q^{-1}\boldsymbol{b}_i\right)
       \right],
\end{aligned}
\label{eq:likelihood_appendix}
\end{equation*}
where $h_0(\cdot)=\boldsymbol{\gamma}_{\mathrm{bs}}^{\top}\boldsymbol{x}_{\mathrm{Spline}}(\cdot)$.
\vspace{0.5 cm}

\noindent The expected log-likelihood is
\begin{equation*}
    \mathbb{E} \left[\ell(\boldsymbol{\theta}\mid\boldsymbol{T},\boldsymbol{\delta},
    \boldsymbol{y},\boldsymbol{b})\right]
    = \sum_i\int \ell_i\times p(\boldsymbol b_i\mid T_i,\delta_i,\boldsymbol y_i)\,d \boldsymbol b_i.
\end{equation*}

\subsection*{Model Parameters}

We have to optimize the following parameters:
\[
    \boldsymbol{\beta}\in\mathbb{R}^{A},
    \qquad
    Q = \begin{pmatrix} q_{11} & q_{12} \\ q_{12} & q_{22} \end{pmatrix}
      \in\mathbb{R}^{2\times 2},
    \qquad
    \sigma^2\in\mathbb{R}^{+},
\]
and
\[
    \boldsymbol{\gamma}\in\mathbb{R}^{J},
    \qquad
    \alpha\in\mathbb{R},
    \qquad
    \boldsymbol{\gamma}_{\mathrm{bs}}\in\mathbb{R}^{W}.
\]
For notational reasons, we now construct two separate parameter vectors, one for each sub-model. Considering $ \boldsymbol{\beta}, q_{11}, q_{12}, q_{22}$ and $\sigma^2$, we get a parameter vector for the longitudinal part of the model  $\boldsymbol{\theta}_{\mathrm{long}}\in\mathbb{R}^{g=A+4}$ and for the survival model we get $\boldsymbol{\theta}_{\mathrm{surv}}\in\mathbb{R}^{h=J+1+W}$.

\noindent The full parameter vector hence is
\[
    \boldsymbol{\theta}
    = \bigl(\underbrace{\boldsymbol{\beta},\,\sigma^2,\,q_{11},\,q_{12},\,q_{22}}_{\boldsymbol{\theta}_{\mathrm{long}}},\;
             \underbrace{\boldsymbol{\gamma},\,\alpha,\,\boldsymbol{\gamma}_{\mathrm{bs}}}_{\boldsymbol{\theta}_{\mathrm{surv}}}\bigr).
\]

\subsection*{Score Functions}
As the estimation procedure of a joint model involves a one-step Newton update for the fixed effects $\boldsymbol{\beta}$ in the longitudinal submodel, the uncorrectd score function as well as the Fisher Information are needed. The respective entries of the score vector can be derived via

\paragraph{Derivative with respect to $\beta_a, a=1,\ldots,A$}

\begin{equation*}
\begin{aligned}
    S(\beta_a)
    &= \frac{\partial}{\partial\beta_a}\mathbb{E} \left[\ell(\boldsymbol{\theta}\mid\boldsymbol{T},\boldsymbol{\delta},
    \boldsymbol{y},\boldsymbol{b})\right] \\
    &= \sum_i ~\boldsymbol{x}_{ia}^{\top}
       \frac{\boldsymbol{y}_i-\boldsymbol{x}_i^{\top}\boldsymbol{\beta}
             -\boldsymbol{z}_i^{\top}\mathbb{E}(\boldsymbol{b}_i)}{\sigma^2}
       +\alpha\,\delta_i\,x_{ia}(T_i) \\
    &\quad-\int \left[
         \int_0^{T_i}h_0(s)\,\alpha\,x_{ia}(s)
         \exp\bigl(\boldsymbol{\gamma}^{\top}\boldsymbol{x}_{i}
         +\alpha\,m_i(s)\bigr)\,ds
       \right]
       p\bigl(\boldsymbol b_i\mid T_i,\delta_i,\boldsymbol{y}_{i};\boldsymbol{\theta}\bigr)\,d \boldsymbol b_i.
\end{aligned}
\end{equation*}

\noindent In order to perfom a Fisher Scoring updating scheme using the Firth-corrected score function for the coefficients in the survival submodel $\boldsymbol \theta_{\text{surv}}$, the uncorrected score functions are needed.

\paragraph{Derivative with respect to $\gamma_j$, $j=1,\ldots,J$}

\begin{equation*}
\begin{aligned}
    S(\gamma_j)
    &= \frac{\partial}{\partial\gamma_j}\mathbb{E} \left[\ell(\boldsymbol{\theta}\mid\boldsymbol{T},\boldsymbol{\delta},
    \boldsymbol{y},\boldsymbol{b})\right] \\
    &= \sum_i ~x_{ij} \left(
         \delta_i
         -\int \left[\int_0^{T_i}h_0(s)
         \exp \bigl(\boldsymbol{\gamma}^{\top}\boldsymbol{x}_{i}
         +\alpha\,m_i(s)\bigr)\,ds\right]
         p \bigl( \boldsymbol b_i\mid T_i,\delta_i,\boldsymbol{y}_{i};\boldsymbol{\theta}\bigr)\,d \boldsymbol b_i
       \right), \\[4pt]
    S^*(\gamma_j)
    &= S(\gamma_j)
       +\frac{1}{2}\,\mathrm{tr} \left(
         I(\boldsymbol{\theta})^{-1}
         \frac{\partial I(\boldsymbol{\theta})}{\partial\gamma_j}
       \right).
\end{aligned}
\end{equation*}

\paragraph{Derivative with respect to $\alpha$}

\begin{equation*}
\begin{aligned}
    S(\alpha)
    &= \frac{\partial}{\partial\alpha}\mathbb{E} \left[\ell(\boldsymbol{\theta}\mid\boldsymbol{T},\boldsymbol{\delta},
    \boldsymbol{y},\boldsymbol{b})\right] \\
    &= \sum_i ~m_i(T_i)\,\delta_i
       -\int \left[
         \int_0^{T_i}h_0(s)\,m_i(s)
         \exp \bigl(\boldsymbol{\gamma}^{\top}\boldsymbol{x}_{i}
         +\alpha\,m_i(s)\bigr)\,ds
       \right]
       p \bigl(\boldsymbol b_i\mid T_i,\delta_i,\boldsymbol{y}_{i};\boldsymbol{\theta}\bigr)\,d \boldsymbol b_i, \\[4pt]
    S^*(\alpha)
    &= S(\alpha)
       +\frac{1}{2}\,\mathrm{tr} \left(
         I(\boldsymbol{\theta})^{-1}
         \frac{\partial I(\boldsymbol{\theta})}{\partial\alpha}
       \right).
\end{aligned}
\end{equation*}

\paragraph{Derivative with respect to $\gamma_{\mathrm{bs}_w}$, $w=1,\ldots,W$}

\begin{equation*}
\begin{aligned}
    S(\gamma_{\mathrm{bs}_w})
    &= \frac{\partial}{\partial\gamma_{\mathrm{bs}_w}}\mathbb{E} \left[\ell(\boldsymbol{\theta}\mid\boldsymbol{T},\boldsymbol{\delta},
    \boldsymbol{y},\boldsymbol{b})\right] \\
    &= \sum_i ~x_{\mathrm{Spline}_{iw}}(T_i)\,\delta_i
       -\int \left[
         \int_0^{T_i}x_{\mathrm{Spline}_{iw}}(s)
         \exp \bigl(\boldsymbol{\gamma}^{\top}\boldsymbol{x}_{i}
         +\alpha\,m_i(s)\bigr)\,ds
       \right]
       p \bigl(b_i\mid T_i,\delta_i,\boldsymbol{y}_{i};\boldsymbol{\theta}\bigr)\,db_i, \\[4pt]
    S^*(\gamma_{\mathrm{bs}_w})
    &= S(\gamma_{\mathrm{bs}_w})
       +\frac{1}{2}\,\mathrm{tr} \left(
         I(\boldsymbol{\theta})^{-1}
         \frac{\partial I(\boldsymbol{\theta})}{\partial\gamma_{\mathrm{bs}_w}}
       \right).
\end{aligned}
\end{equation*}

\newpage
\subsection*{Fisher Information Matrix (Second Derivatives)}

For the one step Newton updating step for the fixed effects regression coefficients $\boldsymbol{\beta}$, the elements of the Hessian $H(\beta_a,\beta_p)$ are used and can be calculated via
\footnotesize{
\begin{equation*}
\begin{aligned}
    -I(\beta_a,\beta_p) = H(\beta_a,\beta_p)
    &= \sum_i ~\frac{\boldsymbol{x}_{ia}^{\top}\boldsymbol{x}_{ip}}{\sigma^2}
       -\int \left[ \int_0^{T_i}h_0(s)\,\alpha^2\,x_{ia}(s)\,x_{ip}(s)
         \exp \bigl(\boldsymbol{\gamma}^{\top}\boldsymbol{x}_{i}
         +\alpha\,m_i(s)\bigr)\,ds\right]
       p \bigl(b_i\mid T_i,\delta_i,\boldsymbol{y}_{i};\boldsymbol{\theta}\bigr)\,db_i, \\[6pt]
\end{aligned}
\end{equation*}
}

\noindent Regarding the survival coefficients $\boldsymbol \theta_{\text{surv}}$, the Fisher-Information matrix is needed for the correction term involved in the Firth-corrected score function and the respective elements are obtained via

\footnotesize{
\begin{equation*}
\begin{aligned}
    -I(\gamma_j,\gamma_k) = H(\gamma_j,\gamma_k)
    &= \sum_i ~-\int \left[\int_0^{T_i}h_0(s)\exp \bigl(\boldsymbol{\gamma}^{\top}\boldsymbol{x}_{i}
         +\alpha\,m_i(s)\bigr)\,x_{ij}\,x_{ik}\,ds\right]
       p \bigl(b_i\mid T_i,\delta_i,\boldsymbol{y}_{i};\boldsymbol{\theta}\bigr)\,db_i, \\[6pt]
    -I(\gamma_j,\alpha) = H(\gamma_j,\alpha)
    &= \sum_i ~-\int \left[\int_0^{T_i}h_0(s)\,m_i(s)
         \exp \bigl(\boldsymbol{\gamma}^{\top}\boldsymbol{x}_{i}
         +\alpha\,m_i(s)\bigr)\,x_{ij}\,ds\right]
       p \bigl(b_i\mid T_i,\delta_i,\boldsymbol{y}_{i};\boldsymbol{\theta}\bigr)\,db_i, \\[6pt]
    -I(\gamma_j,\gamma_{\mathrm{bs}_w}) = H(\gamma_j,\gamma_{\mathrm{bs}_w})
    &= \sum_i ~-\int \left[\int_0^{T_i}x_{\mathrm{spline}_{iw}}(s)
         \exp \bigl(\boldsymbol{\gamma}^{\top}\boldsymbol{x}_{i}
         +\alpha\,m_i(s)\bigr)\,x_{ij}\,ds\right]
       p \bigl(b_i\mid T_i,\delta_i,\boldsymbol{y}_{i};\boldsymbol{\theta}\bigr)\,db_i, \\[6pt]
    -I(\alpha,\alpha) = H(\alpha,\alpha)
    &= \sum_i ~-\int \left[\int_0^{T_i}h_0(s)\,m_i(s)^2
         \exp \bigl(\boldsymbol{\gamma}^{\top}\boldsymbol{x}_{i}
         +\alpha\,m_i(s)\bigr)\,ds\right]
       p \bigl(b_i\mid T_i,\delta_i,\boldsymbol{y}_{i};\boldsymbol{\theta}\bigr)\,db_i, \\[6pt]
    -I(\alpha,\gamma_{\mathrm{bs}_w}) = H(\alpha,\gamma_{\mathrm{bs}_w})
    &= \sum_i ~-\int \left[\int_0^{T_i}x_{\mathrm{spline}_{iw}}(s)\,m_i(s)
         \exp \bigl(\boldsymbol{\gamma}^{\top}\boldsymbol{x}_{i}
         +\alpha\,m_i(s)\bigr)\,ds\right]
       p \bigl(b_i\mid T_i,\delta_i,\boldsymbol{y}_{i};\boldsymbol{\theta}\bigr)\,db_i, \\[6pt]
    -I(\gamma_{\mathrm{bs}_w},\gamma_{\mathrm{bs}_u}) = H(\gamma_{\mathrm{bs}_w},\gamma_{\mathrm{bs}_u})
    &= \sum_i ~-\int \left[\int_0^{T_i}x_{\mathrm{spline}_{iw}}(s)\,x_{\mathrm{spline}_{iu}}(s)
         \exp \bigl(\boldsymbol{\gamma}^{\top}\boldsymbol{x}_{i}
         +\alpha\,m_i(s)\bigr)\,ds\right]
       p \bigl(b_i\mid T_i,\delta_i,\boldsymbol{y}_{i};\boldsymbol{\theta}\bigr)\,db_i.
\end{aligned}
\end{equation*}
}
\subsection*{Derivatives of the Fisher Information Matrix (Third Derivatives)}

The last component for the Firth-correction term, are the third derivatives of the log-likelihood with respect to each element in $\boldsymbol \theta_{\text{surv}}$:

\begin{equation*}
\begin{aligned}
    -\frac{\partial H(\,\cdot\,,\,\cdot\,)}{\partial\gamma_l}
    &= -x_{il}\,H(\,\cdot\,,\,\cdot\,), \\[4pt]
    -\frac{\partial H(\,\cdot\,,\,\cdot\,)}{\partial\alpha}
    &= -m_i(s)\,H(\,\cdot\,,\,\cdot\,), \\[4pt]
    -\frac{\partial H(\,\cdot\,,\,\cdot\,)}{\partial\gamma_{\mathrm{bs}_v}}
    &= -x_{\mathrm{Spline}_{iv}}(s)\,H(\,\cdot\,,\,\cdot\,),
\end{aligned}
\end{equation*}
where the derivatives with respect to $\alpha$ and $\gamma_{\mathrm{bs}_w}$ are understood to be
included inside the time integral $\int_0^{T_i}\cdot\,ds$.

\subsection*{Analytical Solutions for $\hat{\sigma}^2$ and $\hat{\boldsymbol{Q}}$}

The analytical derivatives follow \citet[][Appendix~B]{rizbook}. The closed-form M-step updates require the posterior mean and variance of the random effects,
\begin{equation*}
\begin{aligned}
    \tilde{\boldsymbol{b}}_i
    = \mathbb{E}(\boldsymbol{b}_i)
    &= \int\boldsymbol{b}_i\,
         p\bigl(\boldsymbol{b}_i\mid T_i,\delta_i,\boldsymbol{y}_{i};\boldsymbol{\theta}\bigr)
         \,d\boldsymbol{b}_i, \\[4pt]
    p\bigl(\boldsymbol{b}_i\mid T_i,\delta_i,\boldsymbol{y}_{i};\boldsymbol{\theta}\bigr)
    &= \frac{p(\boldsymbol{y}_i,T_i,\delta_i,\boldsymbol{b}_i)}
            {p(\boldsymbol{y}_i,T_i,\delta_i)}, \\[4pt]
    p(\boldsymbol{y}_i,T_i,\delta_i,\boldsymbol{b}_i)
    &= \exp \Bigl[
         \log p(\boldsymbol{y}_i\mid\boldsymbol{b}_i)
         +\log p(T_i,\delta_i\mid\boldsymbol{b}_i)
         +\log p(\boldsymbol{b}_i)
       \Bigr], \\[4pt]
    p(\boldsymbol{y}_i,T_i,\delta_i)
    &= \int p(\boldsymbol{y}_i,T_i,\delta_i,\boldsymbol{b}_i)\,d\boldsymbol{b}_i,
\end{aligned}
\end{equation*}
and the posterior variance
\begin{equation*}
    \widetilde{\boldsymbol{vb}}_i
    = \mathrm{Var} \left[p \bigl(\boldsymbol{b}_i\mid T_i,\delta_i,\boldsymbol{y}_{i};\boldsymbol{\theta}\bigr)\right]
    = \int\bigl(\boldsymbol{b}_i-\tilde{\boldsymbol{b}}_i\bigr)^2\,
         p \bigl(\boldsymbol{b}_i\mid T_i,\delta_i,\boldsymbol{y}_{i};\boldsymbol{\theta}\bigr)
         \,d\boldsymbol{b}_i.
\end{equation*}

\paragraph{Analytical solution for $\hat{\sigma}^2$}

\begin{equation*}
\begin{aligned}
    \hat{\sigma}^2
    &= \frac{1}{D}\sum_i\int
       \bigl(\boldsymbol{y}_i-\boldsymbol{x}_i^{\top}\boldsymbol{\beta}
             -\boldsymbol{z}_i^{\top}\boldsymbol{b}_i\bigr)^{\top}
       \bigl(\boldsymbol{y}_i-\boldsymbol{x}_i^{\top}\boldsymbol{\beta}
             -\boldsymbol{z}_i^{\top}\boldsymbol{b}_i\bigr)
       \times p \bigl(\boldsymbol{b}_i\mid T_i,\delta_i,\boldsymbol{y}_{i};\boldsymbol{\theta}\bigr)
       \,d\boldsymbol{b}_i \\[4pt]
    &= \frac{1}{D}\sum_i\left[
         \bigl(\boldsymbol{y}_i-\boldsymbol{x}_i^{\top}\boldsymbol{\beta}\bigr)^{\top}
         \bigl(\boldsymbol{y}_i-\boldsymbol{x}_i^{\top}\boldsymbol{\beta}
               -2\boldsymbol{z}_i^{\top}\tilde{\boldsymbol{b}}_i\bigr)
         +\mathrm{tr} \bigl(\boldsymbol{z}_i^{\top}\boldsymbol{z}_i\,
           \widetilde{\boldsymbol{vb}}_i\bigr)
         +\tilde{\boldsymbol{b}}_i^{\top}\boldsymbol{z}_i^{\top}\boldsymbol{z}_i\,
           \tilde{\boldsymbol{b}}_i
       \right].
\end{aligned}
\end{equation*}

where $D$ is the size of the longitudinal data set.

\paragraph{Analytical solution for $\hat{\boldsymbol{Q}}$}

\begin{equation*}
    \hat{\boldsymbol{Q}}
    = \frac{1}{N}\sum_i\Bigl(
        \widetilde{\boldsymbol{vb}}_i
        +\tilde{\boldsymbol{b}}_i\tilde{\boldsymbol{b}}_i^{\top}
      \Bigr).
\end{equation*}

}

\newpage
\section{Simulation results}
\label{sec:AppendixB}

\begin{table}[H]
\label{tab:sim_res}
\caption{Median Squared Error comparison of joint models}
\centering
\begin{tabular}[t]{rll|r|rrrrrrr}
\toprule
\multicolumn{4}{c}{ } & \multicolumn{3}{c}{Survival coefficients} & \multicolumn{4}{c}{Longitudinal coefficients} \\
\cmidrule(l{3pt}r{3pt}){5-7} \cmidrule(l{3pt}r{3pt}){8-11}
N & \%e & separation & method & $\hat \gamma_1$ & $\hat \gamma_2$ & $\hat \alpha$ & $\hat \beta_1$ & $\hat \beta_2$ & $\hat \beta_3$ & $\hat \beta_t$\\
\midrule
 &  &  & JM & 7.5213 & 0.2753 & 0.1563 & 0.0148 & 0.0359 & 0.0017 & 0.1027\\
 &  &  & JMc & 7.4333 & 0.2663 & 0.1569 & 0.0148 & 0.0353 & 0.0023 & 0.1020\\
 &  & \multirow{-3}{*}{\raggedright\arraybackslash no sep.} & FC-JM & 9.4244 & 0.2267 & 0.1437 & 0.0127 & 0.0304 & 0.0019 & 0.1162\\
\cmidrule{3-11}
 &  &  & JM & 255.2093 & 0.3760 & 0.3331 & 0.0107 & 0.0452 & 0.0014 & 0.0863\\
 &  &  & JMc & 6.6125 & 0.3365 & 0.3178 & 0.0115 & 0.0411 & 0.0017 & 0.1304\\
 & \multirow{-6}{*}[0.5\dimexpr\aboverulesep+\belowrulesep+\cmidrulewidth]{\raggedright\arraybackslash 10} & \multirow{-3}{*}{\raggedright\arraybackslash sep.} & FC-JM & 6.0938 & 0.3500 & 0.3617 & 0.0113 & 0.0459 & 0.0017 & 0.0995\\
\cmidrule{2-11}
 &  &  & JM & 2.4858 & 0.0614 & 0.0969 & 0.0133 & 0.0385 & 0.0016 & 0.0487\\
 &  &  & JMc & 2.5538 & 0.0675 & 0.0967 & 0.0173 & 0.0389 & 0.0017 & 0.0536\\

 &  & \multirow{-3}{*}{\raggedright\arraybackslash no sep.} & FC-JM & 3.8202 & 0.0716 & 0.0970 & 0.0152 & 0.0379 & 0.0016 & 0.0484\\
\cmidrule{3-11}
 &  &  & JM & 267.5037 & 0.0847 & 0.0645 & 0.0114 & 0.0869 & 0.0016 & 0.0771\\
 &  &  & JMc & 13.5058 & 0.0783 & 0.0620 & 0.0107 & 0.0700 & 0.0024 & 0.0461\\
 & \multirow{-6}{*}[0.5\dimexpr\aboverulesep+\belowrulesep+\cmidrulewidth]{\raggedright\arraybackslash 25} & \multirow{-3}{*}{\raggedright\arraybackslash sep.} & FC-JM & 1.3330 & 0.0671 & 0.0598 & 0.0093 & 0.0691 & 0.0016 & 0.0716\\
\cmidrule{2-11}
 &  &  & JM & 1.1895 & 0.0265 & 0.0285 & 0.0131 & 0.0400 & 0.0020 & 0.0409\\
 &  &  & JMc & 1.1673 & 0.0258 & 0.0270 & 0.0125 & 0.0364 & 0.0020 & 0.0348\\

 &  & \multirow{-3}{*}{\raggedright\arraybackslash no sep.} & FC-JM & 2.1198 & 0.0294 & 0.0289 & 0.0133 & 0.0399 & 0.0020 & 0.0367\\
\cmidrule{3-11}
 &  &  & JM & 267.1746 & 0.0469 & 0.0263 & 0.0149 & 0.0457 & 0.0026 & 0.0623\\
 &  &  & JMc & 13.5255 & 0.0495 & 0.0252 & 0.0155 & 0.0441 & 0.0020 & 0.0489\\

\multirow{-18}{*}[2.5\dimexpr\aboverulesep+\belowrulesep+\cmidrulewidth]{\raggedleft\arraybackslash 50} & \multirow{-6}{*}[0.5\dimexpr\aboverulesep+\belowrulesep+\cmidrulewidth]{\raggedright\arraybackslash 45} & \multirow{-3}{*}{\raggedright\arraybackslash sep.} & FC-JM & 0.4029 & 0.0473 & 0.0270 & 0.0150 & 0.0458 & 0.0026 & 0.0623\\
\cmidrule{1-11}
 &  &  & JM & 1.7947 & 0.0421 & 0.0342 & 0.0057 & 0.0190 & 0.0005 & 0.0436\\
 &  &  & JMc & 1.7951 & 0.0410 & 0.0290 & 0.0060 & 0.0167 & 0.0005 & 0.0418\\
 &  & \multirow{-3}{*}{\raggedright\arraybackslash no sep.} & FC-JM & 2.5909 & 0.0460 & 0.0393 & 0.0051 & 0.0192 & 0.0004 & 0.0383\\
\cmidrule{3-11}
 &  &  & JM & 252.9465 & 0.0401 & 0.0375 & 0.0056 & 0.0174 & 0.0006 & 0.0398\\
 &  &  & JMc & 12.2741 & 0.0403 & 0.0397 & 0.0053 & 0.0166 & 0.0006 & 0.0377\\
 & \multirow{-6}{*}[0.5\dimexpr\aboverulesep+\belowrulesep+\cmidrulewidth]{\raggedright\arraybackslash 10} & \multirow{-3}{*}{\raggedright\arraybackslash sep.} & FC-JM & 0.4054 & 0.0391 & 0.0361 & 0.0054 & 0.0172 & 0.0005 & 0.0392\\
\cmidrule{2-11}
 &  &  & JM & 0.2303 & 0.0123 & 0.0093 & 0.0036 & 0.0223 & 0.0004 & 0.0444\\
 &  &  & JMc & 0.2353 & 0.0133 & 0.0094 & 0.0046 & 0.0204 & 0.0005 & 0.0466\\
 &  & \multirow{-3}{*}{\raggedright\arraybackslash no sep.} & FC-JM & 0.7075 & 0.0126 & 0.0086 & 0.0041 & 0.0252 & 0.0004 & 0.0472\\
\cmidrule{3-11}
 &  &  & JM & 262.7107 & 0.0148 & 0.0167 & 0.0043 & 0.0190 & 0.0005 & 0.0340\\
 &  &  & JMc & 13.6920 & 0.0147 & 0.0145 & 0.0056 & 0.0148 & 0.0005 & 0.0316\\
 & \multirow{-6}{*}[0.5\dimexpr\aboverulesep+\belowrulesep+\cmidrulewidth]{\raggedright\arraybackslash 25} & \multirow{-3}{*}{\raggedright\arraybackslash sep.} & FC-JM & 0.2095 & 0.0143 & 0.0177 & 0.0043 & 0.0190 & 0.0005 & 0.0325\\
\cmidrule{2-11}
 &  &  & JM & 0.2292 & 0.0059 & 0.0046 & 0.0119 & 0.0161 & 0.0007 & 0.0390\\

 &  &  & JMc & 0.2255 & 0.0059 & 0.0046 & 0.0129 & 0.0167 & 0.0007 & 0.0466\\
 &  & \multirow{-3}{*}{\raggedright\arraybackslash no sep.} & FC-JM & 0.1546 & 0.0060 & 0.0048 & 0.0119 & 0.0161 & 0.0006 & 0.0391\\
\cmidrule{3-11}
 &  &  & JM & 243.5933 & 0.0039 & 0.0045 & 0.0146 & 0.0195 & 0.0007 & 0.0345\\
 &  &  & JMc & 15.0159 & 0.0042 & 0.0051 & 0.0131 & 0.0193 & 0.0007 & 0.0345\\
\multirow{-18}{*}[2.5\dimexpr\aboverulesep+\belowrulesep+\cmidrulewidth]{\raggedleft\arraybackslash 250} & \multirow{-6}{*}[0.5\dimexpr\aboverulesep+\belowrulesep+\cmidrulewidth]{\raggedright\arraybackslash 45} & \multirow{-3}{*}{\raggedright\arraybackslash sep.} & FC-JM & 1.5483 & 0.0038 & 0.0042 & 0.0137 & 0.0195 & 0.0006 & 0.0345\\
\bottomrule
\end{tabular}
\end{table}

\newpage
\section{Data description}
\label{sec:AppendixC}

\subsection{Comparison of samples}

\begin{figure}[ht!]
    \centering
    \includegraphics[width=\linewidth]{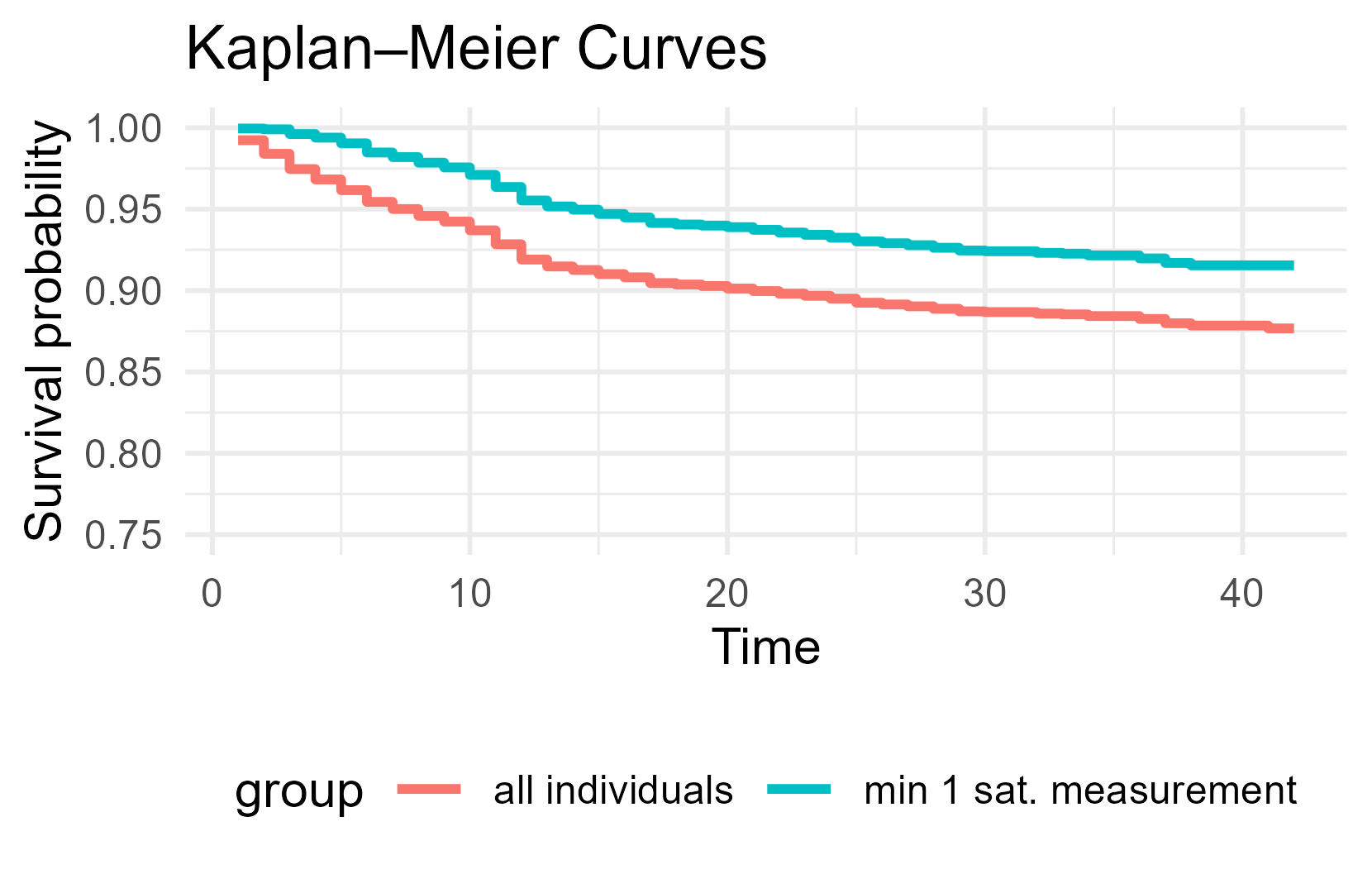}
    \caption{Comparison of Kaplan-Meier-Curves for the complete case sample and the sample with individuals with at least one satisfaction measurement.}
    \label{fig:compare_kmc}
\end{figure}

\begin{sidewaystable}[htbp]
\centering
\small
\begin{tabular}{lcccccc}
\toprule
& \multicolumn{2}{c}{Full sample incl.\ missings} & \multicolumn{2}{c}{Complete cases only} & \multicolumn{2}{c}{\parbox{4cm}{\centering Complete cases only and min.\ one satisfaction measurement}}\\
\cmidrule(lr){2-3} \cmidrule(lr){4-5} \cmidrule(lr){6-7}
Drop out 
& no (N=6,514) & yes (N=847)
& no (N=4,131) & yes (N=455)
& no (N=3,928) & yes (N=275) \\
\midrule
\addlinespace

Education \\
\hspace{1em}Lower secondary (Hauptschule)
& 1,529 (26\%) & 276 (39\%)
& 859 (21\%) & 142 (31\%)
& 798 (20\%) & 91 (33\%) \\
\hspace{1em}Intermediate secondary (Realschule)
& 2,922 (50\%) & 322 (45\%)
& 2,115 (51\%) & 218 (48\%)
& 2,014 (51\%) & 135 (49\%) \\
\hspace{1em}Higher secondary (Gymnasium)
& 1,438 (24\%) & 114 (16\%)
& 1,157 (28\%) & 95 (21\%)
& 1,116 (28\%) & 49 (18\%) \\
\hspace{1em}Unknown
& 625 & 135
& 0 & 0
& 0 & 0 \\
\addlinespace

GPA
& 2.70 (2.30, 3.00) & 2.80 (2.40, 3.10)
& 2.70 (2.30, 3.00) & 2.80 (2.30, 3.00)
& 2.70 (2.30, 3.00) & 2.80 (2.40, 3.10) \\
\hspace{1em}Unknown
& 905 & 186
& 0 & 0
& 0 & 0 \\
\addlinespace

Gender \\
\hspace{1em}Male
& 3,475 (53\%) & 426 (50\%)
& 2,097 (51\%) & 224 (49\%)
& 1,989 (51\%) & 143 (52\%) \\
\hspace{1em}Female
& 3,039 (47\%) & 421 (50\%)
& 2,034 (49\%) & 231 (51\%)
& 1,939 (49\%) & 132 (48\%) \\
\addlinespace

Migrational Background \\
\hspace{1em}None
& 4,915 (76\%) & 562 (67\%)
& 3,310 (80\%) & 328 (72\%)
& 3,176 (81\%) & 206 (75\%) \\
\hspace{1em}First generation
& 376 (5.9\%) & 74 (8.9\%)
& 171 (4.1\%) & 32 (7.0\%)
& 158 (4.0\%) & 22 (8.0\%) \\
\hspace{1em}Second generation
& 1,134 (18\%) & 199 (24\%)
& 650 (16\%) & 95 (21\%)
& 594 (15\%) & 47 (17\%) \\
\hspace{1em}Unknown
& 89 & 12
& 0 & 0
& 0 & 0 \\
\addlinespace

Parental socio-economic status
& 48 (29, 60) & 45 (29, 59)
& 50 (31, 62) & 48 (30, 65)
& 50 (31, 62) & 48 (31, 68) \\
\hspace{1em}Unknown
& 794 & 133
& 0 & 0
& 0 & 0 \\
\addlinespace

Type of training \\
\hspace{1em}School-based
& 1,504 (23\%) & 229 (27\%)
& 921 (22\%) & 131 (29\%)
& 810 (21\%) & 88 (32\%) \\
\hspace{1em}Firm-based
& 4,991 (77\%) & 618 (73\%)
& 3,210 (78\%) & 324 (71\%)
& 3,118 (79\%) & 187 (68\%) \\
\hspace{1em}Unknown
& 19 & 0
& 0 & 0
& 0 & 0 \\
\addlinespace

Place of Residence \\
\hspace{1em}West Germany
& 5,649 (87\%) & 711 (84\%)
& 3,609 (87\%) & 391 (86\%)
& 3,422 (87\%) & 242 (88\%) \\
\hspace{1em}East Germany
& 861 (13\%) & 136 (16\%)
& 522 (13\%) & 64 (14\%)
& 506 (13\%) & 33 (12\%) \\
\addlinespace

Compromise in field of work \\
\hspace{1em}None
& 2,190 (41\%) & 187 (28\%)
& 1,749 (42\%) & 125 (27\%)
& 1,702 (43\%) & 70 (25\%) \\
\hspace{1em}Weak
& 976 (18\%) & 127 (19\%)
& 775 (19\%) & 105 (23\%)
& 728 (19\%) & 69 (25\%) \\
\hspace{1em}Moderate
& 786 (15\%) & 108 (16\%)
& 591 (14\%) & 70 (15\%)
& 551 (14\%) & 45 (16\%) \\
\hspace{1em}High
& 1,363 (26\%) & 254 (38\%)
& 1,016 (25\%) & 155 (34\%)
& 947 (24\%) & 91 (33\%) \\
\hspace{1em}Unknown
& 1,199 & 171
& 0 & 0
& 0 & 0 \\
\addlinespace

Sector of voc. training \\
\hspace{1em}Production of goods
& 2,229 (35\%) & 232 (28\%)
& 1,360 (33\%) & 118 (26\%)
& 1,311 (33\%) & 82 (30\%) \\
\hspace{1em}Personal services
& 1,952 (31\%) & 386 (46\%)
& 1,265 (31\%) & 216 (47\%)
& 1,188 (30\%) & 120 (44\%) \\
\hspace{1em}Business administration \& other business related services
& 1,638 (26\%) & 167 (20\%)
& 1,162 (28\%) & 96 (21\%)
& 1,103 (28\%) & 56 (20\%) \\
\hspace{1em}Services in IT and natural sciences
& 260 (4.1\%) & 24 (2.9\%)
& 182 (4.4\%) & 15 (3.3\%)
& 169 (4.3\%) & 12 (4.4\%) \\
\hspace{1em}Other occupations in commercial services
& 285 (4.5\%) & 25 (3.0\%)
& 162 (3.9\%) & 10 (2.2\%)
& 157 (4.0\%) & 5 (1.8\%) \\
\hspace{1em}Unknown
& 150 & 13
& 0 & 0
& 0 & 0 \\
\bottomrule
\multicolumn{7}{l}{\textsuperscript{1} Median (Q1, Q3); n (\%)}\\
\end{tabular}
\end{sidewaystable}

\newpage
\section{Results of JM and FC-JM}
\label{append: jm_vs_fcjm}

\begin{sidewaystable}[htb!]
\centering
\caption{Comparison of estimation results on NEPS data for women: Classical joint model using \texttt{JM()} and the Firth-corrected joint model.}
\begin{tabular}{l|p{6.2cm}rrr|rrr}
\toprule
&& \multicolumn{3}{c}{Classical joint model} & \multicolumn{3}{c}{Firth-corrected joint model} \\
\cmidrule(lr){3-5} \cmidrule(lr){6-8}
Submodel & Covariate & Value & Std.\ err. & p-value & Value & Std.\ err. & p-value \\ 
\midrule
\multirow{17}{2.7cm}{Longitudinal (Satisfaction)} 
& (Intercept) & 8.4950 & 0.2240 & 0.0000 & 8.5135 & 0.2238 & 0.0000 \\
& Time & -0.4029 & 0.0249 & 0.0000 & -0.4022 & 0.0249 & 0.0000 \\
& Parental socioeconomic status (ISEI) & -0.1076 & 0.0323 & 0.0009 & -0.1072 & 0.0322 & 0.0008 \\
& GPA & -0.0524 & 0.0305 & 0.0865 & -0.0519 & 0.0305 & 0.0869 \\
& Residence Region: East Germany $^a$ & -0.0619 & 0.0880 & 0.4818 & -0.0617 & 0.0878 & 0.4728 \\
& Type of training: firm-based $^b$ & 0.0291 & 0.0741 & 0.6941 & 0.0274 & 0.0739 & 0.6969 \\
& Field of work: weak compromise $^c$ & -0.1270 & 0.0835 & 0.1281 & -0.1246 & 0.0833 & 0.1318 \\
& Field of work: moderate compromise $^c$ & -0.1834 & 0.0913 & 0.0446 & -0.1811 & 0.0911 & 0.0459 \\
& Field of work: strong compromise $^c$ & -0.2911 & 0.0786 & 0.0002 & -0.2883 & 0.0785 & 0.0002 \\
& Migration background: first generation $^d$ & -0.0360 & 0.1487 & 0.8087 & -0.0335 & 0.1484 & 0.8051 \\
& Migration background: second generation $^d$ & 0.0074 & 0.0824 & 0.9287 & 0.0070 & 0.0822 & 0.9133 \\
& Education: intermediate secondary $^e$ & -0.0799 & 0.0859 & 0.3526 & -0.0817 & 0.0858 & 0.3339 \\
& Education: higher secondary $^e$ & -0.1764 & 0.0940 & 0.0605 & -0.1800 & 0.0938 & 0.0538 \\
& Sector: production of goods $^f$ & 0.2684 & 0.2220 & 0.2266 & 0.2549 & 0.2217 & 0.2452 \\
& Sector: personal services $^\text{f}$ & 0.2245 & 0.2043 & 0.2718 & 0.2093 & 0.2041 & 0.2991 \\
& Sector: business administration \& other business related services $^\text{f}$ & 0.1567 & 0.2052 & 0.4450 & 0.1412 & 0.2050 & 0.4811 \\
& Sector: other occ. in commercial services $^\text{f}$ & 0.5093 & 0.3081 & 0.0984 & 0.5007 & 0.3076 & 0.1016 \\
\midrule
\multirow{16}{2.7cm}{Time-to-event (Risk of dropout)} 
& Parental socioeconomic status (ISEI) & 0.1237 & 0.0946 & 0.1911 & 0.1255 & 0.0909 & 0.1641 \\
& GPA & 0.1058 & 0.0907 & 0.2434 & 0.1065 & 0.0874 & 0.2185 \\
& Residence Region: East Germany $^a$ & -0.2762 & 0.2781 & 0.3206 & -0.2495 & 0.2662 & 0.3417 \\
& Type of training: firm-based $^b$ & -0.0454 & 0.2071 & 0.8265 & -0.0476 & 0.1998 & 0.7956 \\
& Field of work: weak compromise $^c$ & 1.0707 & 0.2529 & 0.0000 & 1.0607 & 0.2435 & 0.0000 \\
& Field of work: moderate compromise $^c$ & 0.5212 & 0.2984 & 0.0807 & 0.5206 & 0.2868 & 0.0681 \\
& Field of work: strong compromise $^c$ & 0.8090 & 0.2549 & 0.0015 & 0.7909 & 0.2450 & 0.0012 \\
& Migration background: first generation $^d$ & 0.4048 & 0.3577 & 0.2579 & 0.4498 & 0.3419 & 0.1845 \\
& Migration background: second generation $^d$ & -0.2254 & 0.2629 & 0.3913 & -0.2062 & 0.2526 & 0.4061 \\
& Education: intermediate secondary $^e$ & -0.4150 & 0.2241 & 0.0640 & -0.4392 & 0.2162 & 0.0414 \\
& Education: higher secondary $^e$ & -1.2479 & 0.2738 & 0.0000 & -1.2642 & 0.2634 & 0.0000 \\
& Sector: production of goods $^f$ & 1.8234 & 1.3093 & 0.1637 & 1.0242 & 0.8968 & 0.2484 \\
& Sector: personal services $^\text{f}$ & 1.9928 & 1.2842 & 0.1207 & 1.1657 & 0.8625 & 0.1730 \\
& Sector: business administration \& other business related services $^\text{f}$ & 1.4676 & 1.2891 & 0.2549 & 0.6423 & 0.8697 & 0.4510 \\
& Sector: other occ. in commercial services $^\text{f}$ & -13.5608 & 1748.1488 & 0.9938 & -0.1469 & 1.6393 & 0.9100 \\
& Satisfaction ($\hat \alpha$) & -0.6970 & 0.0984 & 0.0000 & -0.7114 & 0.0933 & 0.0000 \\
\bottomrule
\end{tabular}
\small{ \\ Reference categories:
$^a$ West Germany, 
$^b$ school-based,
$^c$ no compromise,
$^d$ no migration background,
$^e$ lower secondary,
$^\text{f}$ Sector: services in IT and natural sciences}
\label{tab:jm_vs_fcjm_female}
\end{sidewaystable}

\begin{sidewaystable}[htb!]
\centering
\caption{Comparison of estimation results on NEPS data for men: Classical joint model using \texttt{JM()} and the Firth-corrected joint model.}
\begin{tabular}{l|p{6.2cm}rrr|rrr}
\toprule
&& \multicolumn{3}{c}{Classical joint model} & \multicolumn{3}{c}{Firth-corrected joint model} \\
\cmidrule(lr){3-5} \cmidrule(lr){6-8}
Submodel & Covariate & Value & Std.\ err. & p-value & Value & Std.\ err. & p-value \\ 
\midrule
\multirow{17}{2.7cm}{Longitudinal (Satisfaction)} 
& (Intercept) & 8.4025 & 0.1489 & 0.0000 & 8.4085 & 0.1487 & 0.0000 \\
& Time & -0.4197 & 0.0226 & 0.0000 & -0.4192 & 0.0226 & 0.0000 \\
& Parental socioeconomic status (ISEI) & -0.0483 & 0.0289 & 0.0942 & -0.0482 & 0.0289 & 0.0931 \\
& GPA & -0.0931 & 0.0284 & 0.0010 & -0.0927 & 0.0284 & 0.0011 \\
& Residence Region: East Germany $^a$ & -0.1837 & 0.0850 & 0.0308 & -0.1828 & 0.0849 & 0.0307 \\
& Type of training: firm-based $^b$ & 0.3008 & 0.0958 & 0.0017 & 0.2979 & 0.0956 & 0.0018 \\
& Field of work: weak compromise $^c$ & -0.1287 & 0.0767 & 0.0936 & -0.1279 & 0.0766 & 0.0933 \\
& Field of work: moderate compromise $^c$ & -0.2160 & 0.0887 & 0.0148 & -0.2149 & 0.0886 & 0.0150 \\
& Field of work: strong compromise $^c$ & -0.2389 & 0.0723 & 0.0010 & -0.2378 & 0.0722 & 0.0010 \\
& Migration background: first generation $^d$ & -0.2629 & 0.1428 & 0.0657 & -0.2596 & 0.1426 & 0.0673 \\
& Migration background: second generation $^d$ & -0.2433 & 0.0825 & 0.0032 & -0.2422 & 0.0824 & 0.0032 \\
& Education: intermediate secondary $^e$ & 0.0558 & 0.0681 & 0.4128 & 0.0542 & 0.0680 & 0.4170 \\
& Education: higher secondary $^e$ & -0.0812 & 0.0885 & 0.3592 & -0.0833 & 0.0884 & 0.3392 \\
& Sector: production of goods $^f$ & 0.2889 & 0.1173 & 0.0138 & 0.2863 & 0.1172 & 0.0142 \\
& Sector: personenbezogene Dienstleistungsberufe $^f$ & 0.0335 & 0.1368 & 0.8069 & 0.0323 & 0.1367 & 0.7967 \\
& Sector: kaufmännische und unternehmensbezogene Dienstleistungsberufe $^f$ & 0.0100 & 0.1281 & 0.9379 & 0.0082 & 0.1279 & 0.9299 \\
& Sector: IT- and natural sciences services $^f$ & 0.0875 & 0.1595 & 0.5835 & 0.0855 & 0.1594 & 0.5799 \\
\midrule
\multirow{16}{2.7cm}{Time-to-event (Risk of dropout)} 
& Parental socioeconomic status (ISEI) & 0.0459 & 0.0892 & 0.6066 & 0.0461 & 0.0860 & 0.5797 \\
& GPA & 0.1436 & 0.0885 & 0.1045 & 0.1439 & 0.0854 & 0.0900 \\
& Residence Region: East Germany $^a$ & -0.0393 & 0.2693 & 0.8839 & -0.0139 & 0.2579 & 0.9379 \\
& Type of training: firm-based $^b$ & -0.5066 & 0.2230 & 0.0231 & -0.5147 & 0.2153 & 0.0165 \\
& Field of work: weak compromise $^c$ & 0.3673 & 0.2515 & 0.1442 & 0.3714 & 0.2427 & 0.1234 \\
& Field of work: moderate compromise $^c$ & 0.5649 & 0.2702 & 0.0366 & 0.5719 & 0.2609 & 0.0278 \\
& Field of work: strong compromise $^c$ & 0.4593 & 0.2306 & 0.0464 & 0.4567 & 0.2222 & 0.0390 \\
& Migration background: first generation $^d$ & 0.7410 & 0.3281 & 0.0239 & 0.7748 & 0.3137 & 0.0132 \\
& Migration background: second generation $^d$ & 0.2918 & 0.2282 & 0.2010 & 0.3025 & 0.2203 & 0.1663 \\
& Education: intermediate secondary $^e$ & -0.4848 & 0.1942 & 0.0125 & -0.4900 & 0.1877 & 0.0089 \\
& Education: higher secondary $^e$ & -1.1364 & 0.2807 & 0.0001 & -1.1281 & 0.2700 & 0.0000 \\
& Sector: production of goods $^f$ & -0.2855 & 0.3534 & 0.4192 & -0.3282 & 0.3378 & 0.3247 \\
& Sector: personenbezogene Dienstleistungsberufe $^f$ & 0.4225 & 0.3610 & 0.2419 & 0.3869 & 0.3454 & 0.2574 \\
& Sector: kaufmännische und unternehmensbezogene Dienstleistungsberufe $^f$ & -0.4177 & 0.3944 & 0.2895 & -0.4440 & 0.3773 & 0.2345 \\
& Sector: IT- and natural sciences services $^f$ & -0.9384 & 0.5667 & 0.0978 & -0.8984 & 0.5329 & 0.0900 \\
& Satisfaction ($\hat \alpha$) & -0.5754 & 0.0808 & 0.0000 & -0.5835 & 0.0772 & 0.0000 \\
\bottomrule
\end{tabular}
\small{ \\ Reference categories:
$^a$ West Germany, 
$^b$ school-based,
$^c$ no compromise,
$^d$ no migration background,
$^e$ lower secondary,
$^\text{f}$ Sector: services in IT and natural sciences}
\label{tab:jm_vs_fcjm_male}
\end{sidewaystable}

\end{document}